\documentclass[preprint,showpacs,preprintnumbers,superscriptaddress,11pt,graphicx]{revtex4}
\usepackage{graphicx}
\usepackage{dcolumn}
\usepackage{bm}
\usepackage{amssymb}

\newcommand{\bea}{\begin{eqnarray}}
\newcommand{\ena}{\end{eqnarray}}
\newcommand{\be}{\begin{equation}}
\newcommand{\ee}{\end{equation}}

\begin{document}

\title{Constraints on standard and non-standard early Universe models from CMB $B$-mode polarization}
\author{Yin-Zhe Ma}
\email{yzm20@cam.ac.uk}
\affiliation{Kavli Institute for Cosmology Cambridge and Institute of Astronomy,
Madingley Road, Cambridge, CB3 0HA, U.K.}
\author{Wen Zhao}
\email{Wen.Zhao@astro.cf.ac.uk}
\affiliation{School of Physics and Astronomy, Cardiff University, Cardiff, CF24 3AA, U.K. }
\affiliation{Wales Institute of Mathematical and Computational Sciences, Swansea, SA2
8PP, U.K. }
\author{Michael L. Brown}
\email{mbrown@ast.cam.ac.uk}
\affiliation{Kavli Institute for Cosmology Cambridge and Astrophysics Group, Cavendish
Laboratory, J. J. Thomson Avenue, Cambridge CB3 0HE, U.K.}
\date{\today}

\begin{abstract}
We investigate the observational signatures of three models of the
early Universe in the $B$-mode polarization of the Cosmic
Microwave Background (CMB) radiation. In addition to the standard
single field inflationary model, we also consider the constraints
obtainable on the loop quantum cosmology model (from Loop Quantum
Gravity) and on cosmic strings, expected to be copiously produced
during the latter stages of Brane inflation.  We first examine the
observational features of the three models, and then use current
$B$-mode polarization data from the BICEP and QUaD experiments to
constrain their parameters. We also examine the detectability of
the primordial $B$-mode signal predicted by these models and
forecast the parameter constraints achievable with future CMB
polarization experiments. We find that: (a) these three models of
the early Universe predict different features in the CMB B-mode
polarization power spectrum, which are potentially distinguishable
from the CMB experiments; (b) since $B$-mode polarization
measurements are mostly unaffected by parameter degeneracies, they
provide the cleanest probe of these early Universe models; (c)
using the BICEP and QUaD data we obtain the following parameter
constraints: $r=0.02^{+0.31}_{-0.26}$ ($1\sigma$ for the
tensor-to-scalar ratio in the single field inflationary model); $m
< 1.36\times 10^{-8} \text{M}_{\text{pl}}$ and $k_{*} < 2.43
\times 10^{-4} \text{Mpc}^{-1}$ ($1\sigma$ for the mass and scale
parameters in the loop quantum cosmology model); and $G\mu < 5.77
\times 10^{-7}$ ($1\sigma$ for the cosmic string tension); (d)
future CMB observations (both satellite missions and forthcoming
sub-orbital experiments) will provide much more rigorous tests of
these early Universe models.
\end{abstract}

\maketitle

\section{Introduction}

Observations of the Cosmic Microwave Background (CMB) radiation have
proved a valuable tool for studying the physics of the very early
Universe. Scalar, vector and tensor perturbations generated in the
early Universe have left observable imprints in the temperature and
polarization anisotropies of the CMB. Recent experiments, including
the Wilkinson Microwave Anisotropy Probe (WMAP) satellite
\cite{Bennett03,Hinshaw07,Komatsu10}, QUaD \cite{quad}, BICEP
\cite{bicep} and others \cite{other1, other2, other3, other4, other5},
have led to a precise determination of the basic parameters of the
standard $\Lambda$CDM cosmological model, including the parameters
describing the primordial density perturbations.

According to this concordance model, the Universe underwent a period
of near-exponential expansion, termed inflation, at very early
times. The standard model of inflation is based on the single field
slow-roll scenario. In this scenario, the expansion is driven by a
scalar field (the inflaton) gradually rolling down a flat potential
during the inflationary stage. Inflation ended when the slow-roll
conditions were broken, and the inflaton decayed into relativistic
particles which re-heated the Universe.

In spite of many phenomenological successes of inflation based on
effective field theory, serious problems remain concerning the origin
of the scalar field driving inflation, namely the singularity problem
\cite{singularity} and the trans-Plankian problem
\cite{transplanck}. Consequently, efforts have been made to realize
inflation in a more natural way from some fundamental theory of
microscopic physics. Brane inflation \cite{kklt,Kachru03} from high
dimensional string theory is a typical example. In this scenario,
the Universe is embedded into a high dimensional warped space-time.
The anti-brane is fixed at the bottom of a warped throat, while the
brane is mobile and experiences a small attractive force towards the
anti-brane. Inflation ends when the brane and the anti-brane collide
and annihilate, initiating the hot big bang epoch. During the brane
collision, cosmic strings would be copiously produced, and would leave
an imprint on the CMB sky \cite{Jones02,Sarangi02}. Searching for this
cosmic string signal in the CMB is an important way to test the
correctness of this scenario.

Another approach for realizing a period of inflation, based
on Loop Quantum Gravity (LQG) has been proposed recently (see
\cite{lqc} for instance). LQG is a non-perturbative and
background-independent quantization of General Relativity. Based
on a canonical approach, it uses Ashtekar variables, namely SU(2)
valued connections and conjugate densitized triads. The
quantization is obtained through holonomies of the connections and
fluxes of the densitized triads. More importantly, when the energy
density of the Universe was approaching the critical density
$\rho_{c}$, the Universe entered into a bouncing period due to
repulsive quantum geometrical effects. Thus, the big bang is
replaced by a ``big bounce''. This, to some extent, avoids the
singularity problem in the standard $\Lambda$CDM model.

Differentiating between these three classes of models (standard
field inflation, brane inflation and loop quantum cosmology),
which are motivated by different microscopic physics, is a
crucially important goal for modern cosmology. Since the
primordial scalar, vector and tensor perturbations produced in
these models are quite different from each other, they will in
general leave different signatures in the CMB radiation. Numerous
authors have constrained the parameters of standard field
inflation models from CMB and large scale structure observations.
(See for example \cite{Komatsu10, quad, zbg} for some recent
analyses.) Most of these analyses have made use of both
temperature and polarization CMB measurements and their
constraints have been dominated by the temperature measurements.
However, with the advent of a new generation of CMB polarization
measurements \cite{bicep, quad}, it is now possible to obtain
meaningful constraints solely from measurements of the $B$-mode
polarization of the CMB. Since $B$-mode polarization on very large
scale is generated only by tensor perturbations in the early
Universe, this is a particularly attractive technique: a detection
of $B$-mode polarization at large enough angular scales
\emph{must} be due to gravitational waves (tensor perturbations)
and the connection with early Universe physics is then very clear.
In addition, small scale $B$-mode polarization is possibly sourced
by cosmic string. This potential of $B$-mode measurements to
constrain inflation parameters was demonstrated recently to great
effect by the BICEP collaboration who used their $B$-mode
constraints on large angular scales to obtain a 95\% upper limit
on the tensor-to-scalar ratio of $r < 0.73$. Note that on small
scales, $B$-modes are also generated by gravitational lensing of
the dominant $E$-mode polarization signal and so in general,
large-angular scale measurements are required in order to avoid
confusion from the signal due to lensing.


In this paper, we extend the investigation of inflationary constraints
from CMB $B$-mode polarization measurements alone by considering the
constraints obtainable on the three classes of models described above.
The paper is organized as follows. In Section~\ref{Bmode}, we briefly
review the characterization of CMB polarization in terms of $E$- and
$B$-modes. In Section~\ref{observations}, we present the currently
available $B$-mode constraints, and the predicted noise levels for some
current and future CMB experiments. In Section~\ref{data-analysis},
we first present the likelihood and hyper-parameter analysis methods,
which we will use in the parameter estimation. We then describe the
Fisher information matrix formalism which we will use to forecast the
constraints obtainable using future experiments.  In
Section~\ref{sfi-power-spectrum}, we discuss the tensor perturbations
which arise in the single field inflationary (hereafter SFI)
model \footnote{%
Here we restrict our discussion to the single field
slow-roll inflationary model. SFI models with non-trivial sound speed
are not covered here. Thanks to discussion with Daniel Baumann.},
and in Section~\ref{constraints-sfi}, we constrain the parameters of
the SFI model using the BICEP and QUaD data. In
Section~\ref{fisher-sfi}, we calculate the single-to-noise ratio for a
number of future CMB experiments and their combinations, and present
forecasts for the constraints obtainable on the tensor-to-scalar ratio
$r$ with future observations. In Section~\ref{sfi-model-discussion},
we discuss four types of single field slow-roll inflation models, and
their detectability with future experiments. We then follow a similar
line of discussion for the LQG model in Section~\ref{lqg-model} and
the brane inflation/cosmic string model in
Section~\ref{cosmic-string}. We summarize our results in
Section~\ref{conclusion}.

\section{$B$-mode polarization and its observations}
\label{Bmode-and-observations}
\subsection{$B$-mode polarization}
\label{Bmode} Let us first briefly review the statistics of the CMB
polarization field. The polarized radiation field can be described by
a $2 \times 2 $ intensity matrix $I_{ij}(\mathbf{n})$ \cite{chan},
where $\mathbf{n}$ denotes the direction on the sky, and
$I_{ij}(\mathbf{n})$ is defined with respect to the orthogonal basis
($\mathbf{e}_{1}$, $\mathbf{e}_{2}$) which is perpendicular to
$\mathbf{n}$. Linear polarization is related to the two Stokes
parameters, $Q=\frac{1}{4}(I_{11}-I_{22})$ and $U=\frac{1}{2}I_{12}$,
whereas the temperature anisotropy is
$T=\frac{1}{4}(I_{11}+I_{22})$. The polarization magnitude and
orientation are given by $P=\sqrt{Q^{2}+U^{2}}$ and
$\alpha=\frac{1}{2}\text{tan}^{-1}(U/Q)$.

As spin $\pm 2$ fields, the Stokes parameters $Q$ and $U$ change under a
rotation by an angle $\psi$ as $(Q\pm iU)(\mathbf{n})\rightarrow e^{\mp
2i\psi}(Q\pm iU)(\mathbf{n})$. Thus, $(Q \pm iU)(\mathbf{n})$ requires an
expansion with spin $\pm 2$ spherical harmonics \cite{zaldarriaga1997}
\begin{equation}
(Q\pm iU)(\mathbf{n})=\sum_{lm}a_{lm}^{(\pm 2)}[_{\pm 2}Y_{lm}(\mathbf{n})] .
\label{QUexpansion}
\end{equation}
The multipole coefficients $a_{lm}^{(\pm 2)}$ can be calculated as
\begin{equation}
a_{lm}^{(\pm 2)} = \int  (Q\pm iU)(\mathbf{n}) [_{\pm 2}Y^*_{lm}(\mathbf{n})] d {\bf n}.
\end{equation}
The $E$- and $B$-mode multipoles are defined in terms of the
coefficients $a_{lm}^{(\pm 2)}$ in the following manner:
\begin{equation}
a_{lm}^{E}=-\frac{1}{2}(a_{lm}^{(2)}+a_{lm}^{(-2)}),\text{ }a_{lm}^{B}=-%
\frac{1}{2i}(a_{lm}^{(2)}-a_{lm}^{(-2)}) .  \label{EBcoefficients}
\end{equation}
One can now define the electric polarization sky map $E(\bf n)$ and
the magnetic polarization sky map $B(\bf n)$ as
\begin{equation}
E(\mathbf{n})=\sum_{lm}a_{lm}^{E}Y_{lm}(\mathbf{n}),\text{ }B(\mathbf{n}%
)=\sum_{lm}a_{lm}^{B}Y_{lm}(\mathbf{n}) .  \label{EBmodes}
\end{equation}


The scalar field $E({\bf n})$ and the pseudoscalar field $B({\bf
n})$ completely describe the polarization field. $E$-modes are
curl-free modes and appear as symmetric radial and tangential
polarization patterns on the sky. $B$-modes are divergence-free modes
with left-handed and right-handed vortical polarization patterns on
the sky.

One constructs the various CMB power spectra by correlating the $T$,
$E$ and $B$ modes in harmonic space. In the absence of
parity-violating effects \footnote{%
Since both $T$ and $E$ have parity factor $(-1)^{l}$ under rotation,
while $B$ has parity $(-1)^{l+1}$, the $TB$ and $EB$ should vanish for
symmetry reasons.},
there are only four non-zero cross-correlations: $TT$,
$TE$, $EE$ and $BB$. The angular power spectra of
the polarization fields are defined as
\begin{eqnarray}
C_{l}^{EE}\equiv \frac{1}{2l+1}\sum_{m} \langle a^{E}_{lm} a^{E*}_{lm}
\rangle,~~ C_{l}^{BB}\equiv \frac{1}{2l+1}\sum_{m} \langle a^{B}_{lm}
a^{B*}_{lm} \rangle,  \label{ebpowerspectrum}
\end{eqnarray}
where the brackets denote an ensemble average. If the fluctuations
are Gaussian distributed, all of the cosmological information is
encoded in the angular power spectra of Eq. (\ref{ebpowerspectrum}).


\subsection{Constraints on the $B$-mode signal}
\label{observations}
\begin{figure}[tbp]
\centerline{\includegraphics[bb=0 0 646
416,width=4.2in]{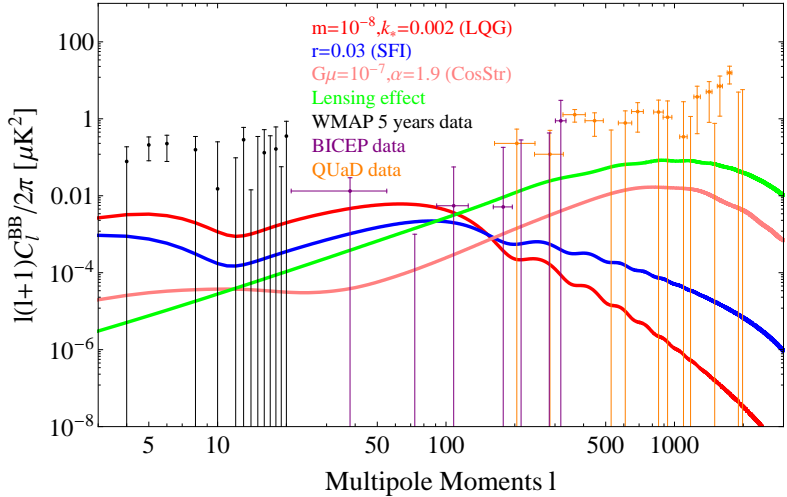}} \caption{Comparison of
different theoretical predictions, and the currently available data
for the $B$-mode power spectrum. The unit for $m$ is $\text{M}_{\text{pl}}$
and $k_{*}$ is $\text{Mpc}^{\text{-1}}$.} \label{datacompare}
\end{figure}
In Fig.~\ref{datacompare}, we plot the current constraints on the
$B$-mode power spectrum along with some representative $B$-mode
signals from different theories. The red curve is the predicted
$C_{l}^{BB}$ calculated from LQG, with a mass parameter
$m=10^{-8} \, \text{M}_{\text{pl}}$ and $k_{*}=0.002 \, \text{Mpc}^{-1}$ (see Section
\ref{perturbation-loop} for a discussion on LQG). The blue curve is
the predicted $C_{l}^{BB}$ from SFI for a tensor-to-scalar ratio
$r=0.03$, and the pink curve is the BB power spectrum generated by
cosmic strings for a string tension $G\mu=10^{-7}$ and wiggling
parameter $\alpha=1.9$ (see Section \ref{cosmic-string} for a
discussion on cosmic strings). The green curve is the $B$-mode signal
from gravitational lensing which acts as a source of confusion when
attempting to measure the primordial $B$-mode signal. The black,
purple and orange points with associated error-bars are the currently
available $B$-mode data from the WMAP 5-year observations ($l \leq 20$
\cite{lambdaweb}), the BICEP experiment (9 band powers
\cite{bicepweb,bicep}), and the QUaD experiment (23 band powers
\cite{quadweb,quad}).

The WMAP constraints are relatively weak due to instrumental noise,
cosmic variance and residual foreground noise. In addition, the
constraining power is further restricted by the uncertainty in the
optical depth $\tau$ to the last scattering surface. We will therefore
not use the WMAP data in the following likelihood analysis. The BICEP
data probes intermediate scales ($21\leq l\leq335 $) around the
recombination bump in the primordial $B$-mode spectrum. On these
scales, the primordial signal is less affected by cosmic variance and
is comparable to or larger than the lensing signal for
tensor-to-scalar ratios $r \gtrsim 0.01$.  The QUaD experiment, whose
primary aim was a high resolution measurement of the $E$-mode signal,
probes small scales ($164\leq l\leq2026$). Its ability to constrain
the primordial signal is thus severely restricted due to lensing
confusion and the rapid decline in the primordial signal with inverse
scale. It may however be useful for constraining the cosmic string
signal which peaks on small scales.

\begin{figure}[tbp]
\centerline{\includegraphics[bb=0 0 594
384,width=3.4in]{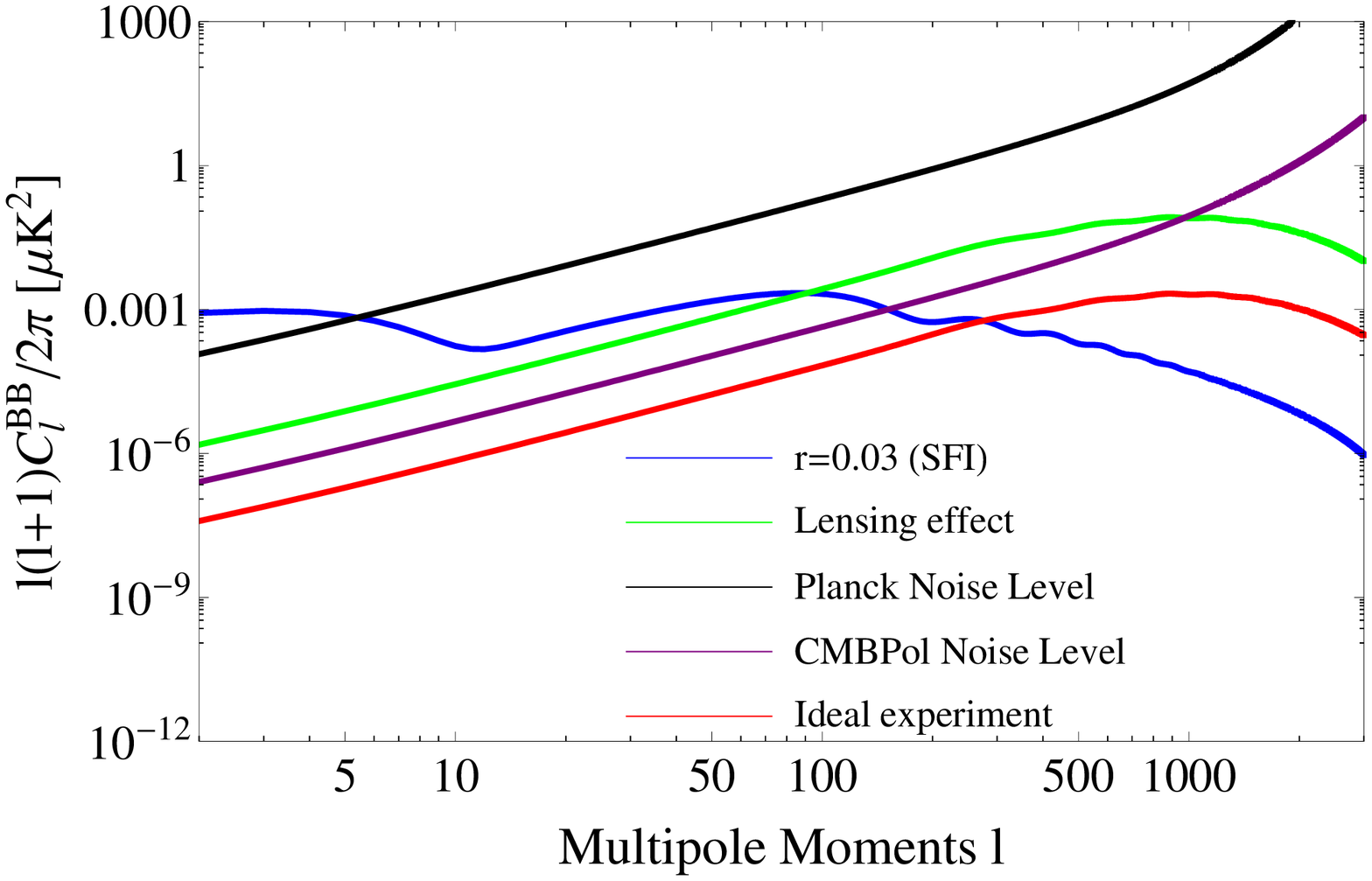}
\includegraphics[bb=0 0 591 381,width=3.4in]{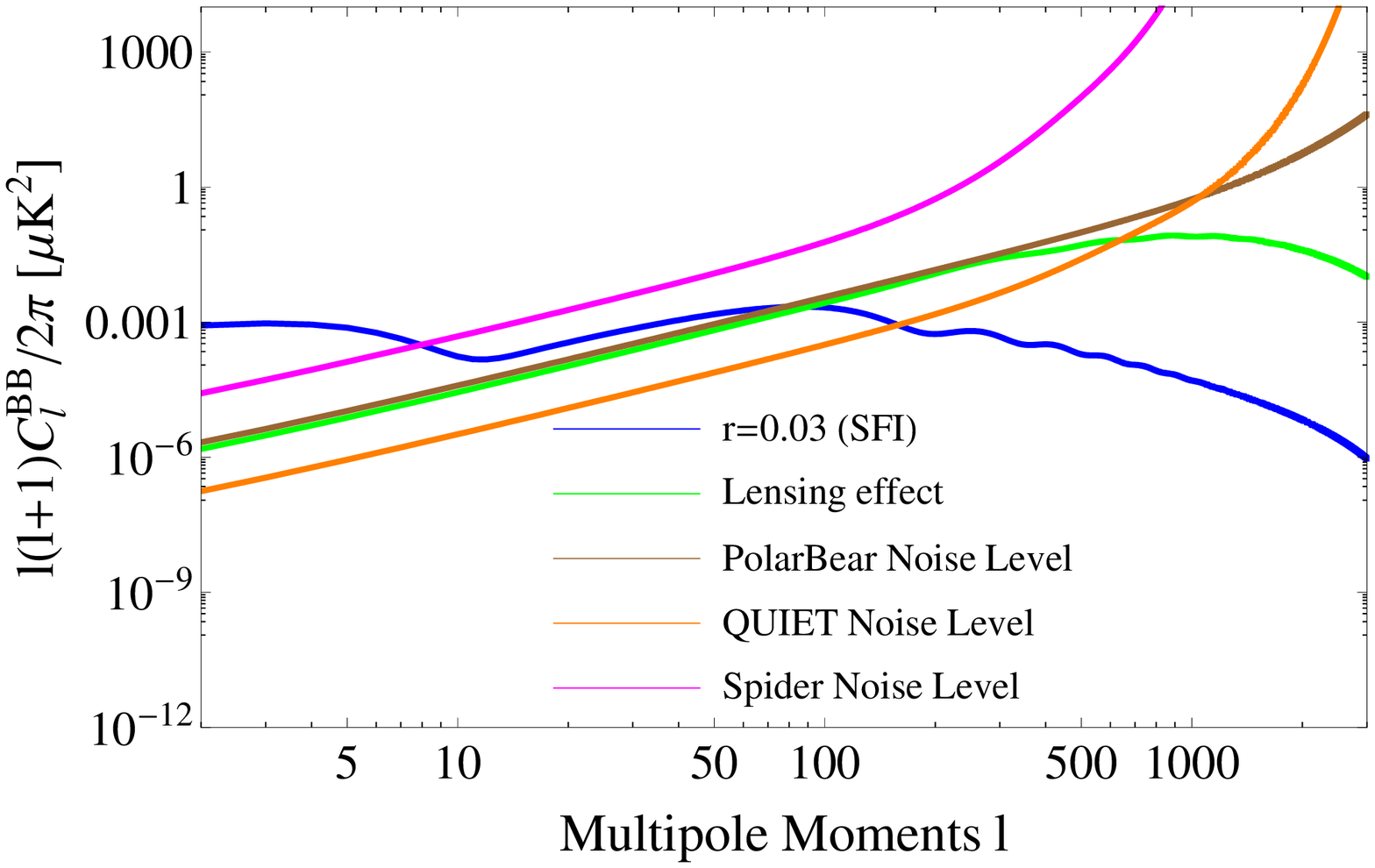}}
\caption{Polarization noise power spectra for forthcoming
  experiments. Note that these curves include uncertainties associated
  with the instrumental beam. The blue curves show the $B$-mode power
  spectrum for the standard inflationary model with $r=0.03$. In the
  left panel, we plot the instrumental noise for the Planck and CMBPol
  satellites, as well as the lensing $B$-mode signal and the noise
  level for the ideal experiment. In the right panel, we plot the
  instrumental noise for the ground-based PolarBear and QUIET
  experiments, and the balloon-borne Spider experiment.}
\label{signalnoise}
\end{figure}

The constraints plotted in Fig. \ref{datacompare} are all consistent
with zero signal at the $\sim 2\sigma$ level. Detecting $B$-mode
polarization therefore remains an outstanding experimental challenge,
and represents a key goal for current and future CMB experiments
including ground-based (BICEP-II \cite{bicep2}, QUIJOTE
\cite{quijote}, PolarBear \cite{polarbear}, QUIET \cite{quiet}),
balloon-borne (EBEX \cite{ebex}, Spider
\cite{spider}, PIPER) and satellite (Planck \cite{planck}, B-Pol \cite{bpol},
litebird \cite{litebird}, CMBPol \cite{cmbpol}) experiments. In what
follows, we will forecast the constraints potentially achievable with
the following five representative experiments: the Planck and CMBPol
satellite missions, the ground-based PolarBear and QUIET (Phase II)
experiments, and the balloon-borne experiment, Spider.  In addition,
for reference, we shall consider the ideal (but unrealistic) case where
there is no foreground contamination and no instrumental noise, and
where the lensing signal can be cleaned to around 1 part in 40
\cite{seljak}. The instrumental specifications which we use to model
the various experiments are listed in Appendix~\ref{instruments}.
Fig.~\ref{signalnoise} shows the noise levels of these experiments
compared to the SFI signal for $r=0.03$.


\subsection{Data Analysis Methodology}
\label{data-analysis} In this subsection, we describe the
methodology we use to constrain the models using current data, and
to forecast constraints for future experiments. The parameters of the
standard $\Lambda$CDM model have already been tightly constrained by
CMB TT, EE and TE data \cite{Komatsu10} and the remaining
uncertainties in these parameters have little impact on the $B$-mode
power spectrum, e.g. \cite{Lewis06}. Therefore, consistent with the
approach adopted by Ref. \cite{bicep}, throughout this paper, we only
vary those parameters which influence the level of primordial
$B$-modes. We fix the other cosmological parameters at their WMAP
7-year best-fit values, which are derived under the assumption $r=0$
and constant $n_s$ across all wavelengths \cite{Komatsu10}:
$\Omega_{b}h^{2}=0.02258$, $\Omega_{c}h^{2}=0.1109$, $n_{s}=0.963$,
$A_{s}(k_{0})=2.43\times10^{-9}$ (pivot scale $k_{0}=0.002 \,
\text{Mpc}^{-1}$), $h=0.71$, and $\tau=0.088$.

\subsubsection{$\chi^{2}$ analysis and hyper-parameters}
\label{likelihood}
To constrain the three models using current data, we initially employ
a conventional $\chi^{2}$ analysis to obtain the likelihood function
for each data set. For LQG and SFI, we use the CAMB code
\cite{CAMBweb} to output the transfer function for the $B$-mode power
spectrum. $C_l^{BB}$ can then be calculated as
\begin{equation}
C_{l}^{BB}=\frac{\pi }{4}\int P_{t}(k)\Delta
_{l}^{B}(k)^{2} d\ln k, \label{clbb-eq}
\end{equation}%
where $\Delta^B_{l}(k)$ is the transfer function for each multipole $l$,
and $P_t(k)$ is the primordial tensor power spectrum (see Eqs.
(\ref{tensorpowlaw}) and (\ref{pt2})). For the cosmic
string model, we use the publicly available code CMBACT \cite{CMBACT}
to generate the $B$-mode power spectrum.

We then follow the pipelines in Refs.~\cite{bicep,quad} to construct
the expected bandpowers for each model, and we use the lognormal
approximation (as illustrated in e.g. \cite{bicep}) to calculate $\chi
^{2}$ according to
\begin{equation}
\chi ^{2}(\alpha )=\left[ \mathbf{\hat{Z}}^{BB}-\mathbf{Z}^{BB}(\alpha )%
\right] ^{T} {\mathbf{D}^{BB}(\alpha)}^{-1} \left[ \mathbf{\hat{Z%
}}^{BB}-\mathbf{Z}^{BB}(\alpha )\right] ,
\end{equation}%
where $\alpha$ is the parameter we wish to constrain, and $\mathbf{Z}%
^{BB}(\alpha )$ and $\hat{\mathbf{Z}}^{BB}$ are the model and
observed band powers, transformed to the lognormal basis.
$\mathbf{D}^{BB}(\alpha)$ is the covariance matrix of the observed
bandpowers, once again transformed to the lognormal basis.
Minimizing the $\chi^{2}$ across all sampled values of $\alpha$
yields the best-fit model.

To obtain joint constraints from BICEP and QUaD, we can simply add the
$\chi^{2}$ values and minimize the resulting joint $\chi^{2}$,
\begin{equation}
\chi^{2}_{\text{tot}}=\chi^{2}_{\text{BICEP}}+\chi^{2}_{\text{QUaD}}. \label{addchi2}
\end{equation}
The goodness of fit for each model can be ascertained by comparing
the minimum $\chi^{2}$ value with the number of degrees of freedom
$n$. If the value of $\chi^{2}_{min}/n$ is close to unity within the
range ($1-\sqrt{2/n}$,$1+\sqrt{2/n}$), we can say that the model
provides a good fit to the data. If $\chi^{2}_{min}/n \gg
1+\sqrt{2/n}$, then the model is not a good fit to the data, while if
$\chi^{2}_{min}/n \ll 1-\sqrt{2/n}$, then the model is overfitting the
data which may happen if the model has redundant free parameters
and/or the errors on the data have been overestimated.

Note that in the conventional joint $\chi^2$ analysis of
Eq.~(\ref{addchi2}), we have weighted each data set equally. This may
be problematic if the two data sets are not mutually consistent, or if
there are unquantified systematics in the data
\cite{Lahav99,Hobson02}. In such cases, one may wish to weight the
data appropriately. The assignment of weights often occurs when two or
more of the data sets are inconsistent, and is usually made in a
somewhat ad-hoc manner \cite{Hobson02}. Generally speaking, assigning
the weights for each data set is a somewhat subjective way of
performing a joint analysis, but one well-motivated approach to
assigning weights is the ``hyper-parameter'' approach, formulated
within a Bayesian context, which can objectively allow the statistical
properties of each data set to determine its own weight in the
analysis \cite{Lahav99,Hobson02}.

In the hyper-parameter technique, the effective $\chi^{2}$ is defined
as
\begin{equation}
\chi _{\text{hyper}}^{2}=\sum_{j}n_{j}\ln \chi _{j}^{2},
\label{hyper-definition}
\end{equation}%
where $j$ sums over all of the data sets, $\chi^{2}_{j}$ is the
$\chi^{2}$ for each data set, and $n_{j}$ is the number of degrees of
freedom for each data set \footnote{There is a subtle difference
between our definition and those in \cite{Lahav99} and
\cite{Hobson02}. In \cite{Lahav99}, $n_{j}$ is the number of data
points in each data set $n_{\text{data}}$, while in
\cite{Hobson02}, $n_{j}=n_{\text{data}}+2$. However, we prove in
Appendix \ref{hyper-parameter} that, when considering several
constraint equations on the random variables, only if the
hyper-parameter is defined as Eq. (\ref{hyper-definition}), the
distribution function form is not changed under the constraints.}.

Once the $\chi^2$ values for the combined data set have been obtained,
we can find the posterior distribution for the parameter $\alpha$
using
\begin{equation}
-2 \ln P(\alpha|D_{1},\cdots, D_{N})=\chi^{2}, \label{likeli1}
\end{equation}
where $\chi^2$ can be either the conventional $\chi^2_{\text{tot}}$ or
the hyper-parameter version $\chi^{2}_{\text{hyper}}$ \cite{Lahav99}.

In Appendix \ref{hyper-parameter}, we calculate the expectation value
and variance (see Eq. (\ref{expectation-hyper})) of
$\chi^2_{\text{hyper}}$. This calculation shows that in the
hyper-parameter case, a model can be said to be a good fit to the data
if the minimum $\chi^{2}$ is within the range $(1-\sqrt{V(n)}/E(n),
1+\sqrt{V(n)}/E(n))$, where $n$ is the number of degrees of freedom
for the data sets \footnote{The Bayesian Evidence (BE) is another
important technique for discriminating between different
models. However, the currently available data is clearly not
constraining enough at present to \emph{discriminate} between
models. We therefore defer any discussion of the BE until more precise
data becomes available.}.

\subsubsection{Fisher information matrix}
\label{fishermatrix} In order to make forecasts for the constraints
achievable with future experiments, one can make use of the Fisher
information matrix under the assumption that each parameter is
Gaussian-distributed. The standard Fisher matrix $F_{\alpha\beta}$ is
defined as \cite{Kendallbook,Tegmark97},
\begin{equation}
F_{\alpha \beta }=\frac{1}{2}\text{Tr}[C_{,\alpha }C^{-1}C_{,\beta }C^{-1}],
\end{equation}%
where $C$ is the total covariance matrix, which includes both signal
and noise contributions:
\begin{equation}
C_{l_{1}m_{1}l_{2}m_{2}}=(C_{l_{1},sig}^{BB}+N_{l_{1},tot}^{BB})\delta
_{l_{1}l_{2}}\delta _{m_{1}m_{2}}.
\end{equation}%
Here, $N_{l,tot}^{BB}$ is the total noise contribution to the
covariance matrix, which includes instrumental noise,
foreground contamination as an effective noise, and confusion noise
from lensing $B$-modes (see Appendix \ref{instruments} for the details).
In our case where we consider only $B$-mode polarization, the Fisher
matrix can be simplified as \cite{Tegmark97,fisher2}
\begin{equation}
F_{\alpha \beta }=\sum_{l}\left( \frac{2l+1}{2}f_{\rm sky}\right) \frac{%
(C_{l,sig}^{BB})_{,\alpha}(C_{l,sig}^{BB})_{,\beta}}{(C_{l,sig}^{BB}+N_{l,tot}^{BB})^{2}},
\label{falphabeta}
\end{equation}%
where $f_{\rm sky}$ is the fraction of sky observed. For Planck,
CMBPol, Spider and the ideal experiment, since these are nearly
full-sky observations, we perform the summation in
Eq. (\ref{falphabeta}) from $l = 2$ to $l = 3000$. For the
ground-based PolarBear and QUIET experiments, the summation is
performed from  $l = 21$ to $l = 3000$. We restrict the summation for
these experiments to $l > 20$ since ground-based experiments are
insensitive to the largest angular scales because of their finite
survey areas (see \cite{jaffe} for instance).

The inverse of the Fisher matrix $F^{-1}$ can, crudely speaking, be
considered the best achievable covariance matrix for the parameters
given the experimental specification. The Cramer-Rao inequality means
that no unbiased method can measure the $i^{th}$ parameter with an
uncertainty (standard deviation) less than $1/\sqrt{F_{ii}}$
\cite{Kendallbook,Tegmark97}. If the other parameters are not known
but are also estimated from the data, the minimum standard deviation
rises to $(F^{-1})^{1/2}_{ii}$ \cite{Kendallbook,Tegmark97}. Therefore
we can estimate the best prospective signal-to-noise ratio as $\alpha/
\Delta \alpha$, where $\Delta \alpha =(F^{-1})_{\alpha \alpha }^{%
  {1}/{2}}$. This formula will be used frequently in the following
discussion.

\section{Constraining the SFI model}
\label{sfi-model}
\subsection{Scalar and tensor primordial power spectra in the SFI model}
\label{sfi-power-spectrum}

In addition to nearly scale-invariant scalar perturbations,
inflationary models also predict vector and tensor perturbations
\cite{grishchuk}. However, the vector perturbations are expected to be
negligible since these modes decayed very rapidly once they entered the
Hubble horizon. We will therefore ignore any vector component in what
follows.

We will work in the perturbed Friedmann-Lemaitre-Robertson-Walker
Universe, for which the metric can be written as
\begin{equation}
ds^2=-c^2dt^2+a^2(t)(\delta_{ij}+h_{ij})dx^idx^j.
\end{equation}
The tensor perturbations $h_{ij}$ are described by two
transverse-traceless components. The power spectrum for the two
polarization modes of $h_{ij}$ ($h_{ij}=h^{+}e_{ij}^{+}+h^{\times}e_{ij}^{\times }$,
$h=h^{+}=h^{\times}$) is
\begin{equation}
\left\langle h_{\mathbf{k}}h_{\mathbf{k}^{\prime }}\right\rangle =(2\pi)^3\delta ^{3}(\mathbf{k}-\mathbf{k}^{\prime })\frac{2\pi ^{2}}{k^{3}}%
P_{t}(k) ,  \label{tensorpower}
\end{equation}
where $h_{\bf k}$ is the Fourier component of the perturbation field
$h$. Standard inflationary models predict a nearly scale-invariant
tensor power spectrum $P_t(k)$. In order to describe the weak
scale-dependence of $P_t(k)$, we can define the tensor spectral index
$n_t$ in the usual way:
\begin{equation}
n_{t}\equiv\frac{d\ln P_{t}(k)}{d\ln k} .  \label{tensortilt}
\end{equation}
The tensor power spectrum can then be written in the following
power-law form
\begin{equation}
P_{t}(k)=A_{t}(k_{0})\left(\frac{k}{k_{0 }}\right)^{n_{t}}
. \label{tensorpowlaw}
\end{equation}

In the case where inflation is driven by a single inflaton field,
the following calculation yields the primordial power spectrum of
the scalar perturbations for slow-roll inflation (see
\cite{Riotto02,Baumann08} for instance. For alternative
calculations, see \cite{grishchuk2007}.)
\begin{eqnarray}
P_{s}(k) &=&\left.\frac{H^{4}}{(2\pi \dot{\phi})^{2}}\right|_{k=aH}  \nonumber \\
&=&\left.\frac{9}{(2\pi )^{2}}\frac{1}{(3\text{M}_{\text{pl}}^{2})^{3}}\frac{V^{3}}{V^{\prime 2}}%
\right|_{k=aH}  \nonumber \\
&=&\frac{8}{3\epsilon }\left( \frac{V^{\frac{1}{4}}}{\sqrt{8\pi }\text{M}_{\text{pl}}}%
\right) ^{4},  \label{ps1}
\end{eqnarray}%
while the power spectrum of tensor perturbations is given by
\begin{eqnarray}
P_{t}(k) &=&\left.\frac{8}{\text{M}_{\text{pl}}^{2}}\left( \frac{H}{2\pi }\right) ^{2}\right|%
_{k=aH}  \nonumber \\
&=&\left.\frac{2}{3}\frac{V}{\pi ^{2}\text{M}_{\text{pl}}^{4}}\right|_{k=aH}.
\label{pt1}
\end{eqnarray}%
Here, $V(\phi)$ is the inflaton potential, and $H$ is the Hubble
parameter at the time of inflation. It is customary to define the
tensor-to-scalar ratio $r$ as
\begin{equation}
r\equiv\frac{P_{t}}{P_{s}}=8\text{M}_{\text{pl}}^{2}\left(
\frac{V^{\prime }}{V}\right) ^{2}.
\label{tsratio2}
\end{equation}%
The tensor-to-scalar ratio and the tensor spectral index are related
to the slow-roll parameter $\epsilon $ via \cite{Riotto02,Baumann08}
\begin{equation}
r=16\epsilon,  \text{   } n_{t}=-2 \epsilon.  \label{tsratio3}
\end{equation}%
These expressions lead to the so-called consistency relation for
single field slow-roll inflation \cite{consistency}:
\begin{equation}
n_{t}=-\frac{r}{8} . \label{consistent-relation}
\end{equation}
Unfortunately, this consistency relation is extremely difficult to
constrain observationally because of the small amplitude of the tensor
power spectrum. We discuss the possibilities for testing this relation
with future observations in Section \ref{fisher-sfi}.

The normalization of the power spectrum of scalar perturbations
(defined at the pivot wavenumber $k_0=0.002$ Mpc$^{-1}$) is
$P_{s}(k_0)=(2.43\pm 0.11)\times 10^{-9}$($1\sigma $ CL, WMAP 7-year
data \cite{Komatsu10}). We can use this normalization together with
Eq. (\ref{ps1}) to derive the relationship between the energy scale of
inflation and the value of $r$:
\begin{equation}
V^{\frac{1}{4}}=1.06\times 10^{16}\text{GeV}\left( \frac{r}{0.01}\right) ^{%
\frac{1}{4}}.  \label{energyscale}
\end{equation}%
That is, a detection of the tensor-to-scalar ratio at $r \approx 0.01$ or
greater would indicate that inflation happened at an energy scale
comparable to the Grand Unification Theory (GUT) energy scale.

We can also use the slow-roll approximation to derive the following
relation, which characterizes the distance in the field space from the
end of inflation to the time when CMB scale fluctuations were
created, namely the Lyth bound \cite{Lyth97,Efstathiou05}
\begin{equation}
\frac{\Delta \phi }{\text{M}_{\text{pl}}}\gtrsim \left(\frac{r}{0.01}\right)^{\frac{1}{2}}
.  \label{lythbound}
\end{equation}
Thus, a tensor-to-scalar ratio greater than $\sim 0.01$, would
directly indicate a super-Planckian field evolving from
$\phi_{CMB}$ to $\phi_{end}$. Such a detection could provide
important observational clues about the nature of quantum gravity.
The boundary $r \sim 0.01$ is therefore an important benchmark
which can confirm or rule out a wide class of large field
inflation models.

\subsection{Constraints on SFI from current data}
\label{constraints-sfi} In this subsection, we present constraints
on the tensor-to-scalar ratio $r$ from BICEP and QUaD data. To
obtain the constraints, we follow the methodology outlined in
Section \ref{likelihood}.
\begin{figure}[tbp]
\centerline{\includegraphics[bb=0 0 614
409,width=4.2in]{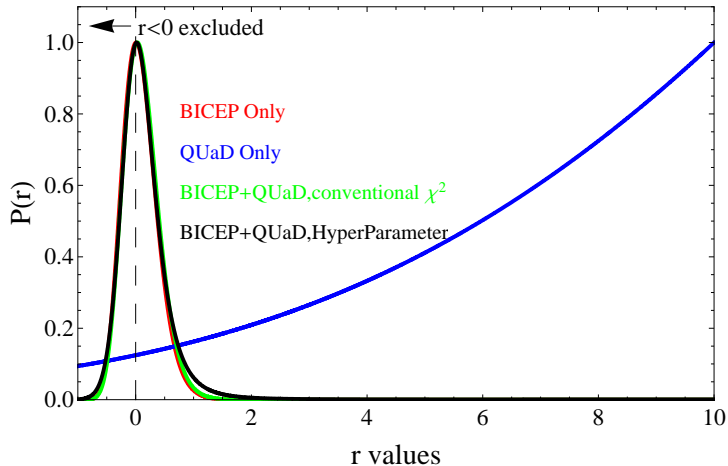}} \caption{The current constraints on
$r$ from BICEP and QUaD data.} \label{rconstrain}
\end{figure}

The results of the analysis are shown in Fig. \ref{rconstrain}. It is
clear from the figure that the BICEP data provides a fairly strong
upper limit on the value of $r$. The best-fit value is
$r=0.01^{+0.31}_{-0.26}$ ($1\sigma$ CL), which is very close to the
result obtained by the BICEP team themselves \cite{bicep}.  As
expected, $r$ is essentially unconstrained by QUaD, whose measurements
are made at much smaller scales ($l \gtrsim 200$) than the scale at
which the primordial $B$-mode signal peaks ($l \sim 80$). In producing
joint constraints, we find that both the conventional $\chi^{2}$ and
the hyper-parameter version ($\chi^{2}_{\text{hyper}}$) are completely
dominated by the BICEP data. For the conventional $\chi^2$ analysis,
the best-fit of the joint analysis gives $r=0.03^{+0.32}_{-0.27}$
($1\sigma$ CL), and for the hyper-parameter $\chi^2$ analysis, we
obtain $r=0.02^{+0.31}_{-0.26}$ ($1\sigma$ CL). The details of the
constraints are listed in the first row of Table \ref{tab1}. Note that
the tendency for the QUaD data to prefer larger values of $r$ appears
to be related to a marginal excess of power in the QUaD measurements
over the multipole range $300 < l < 500$ (see
Fig.~\ref{datacompare}). However, the shape of this apparent
``signal'' is inconsistent with the SFI model (and indeed with any
other primordial $B$-mode model) which suggests that it is likely due
to unquantified residual systematics in the data rather than due to a
true cosmological signal.

Comparing the BICEP result to the WMAP 7-year results
\cite{Komatsu10}, the tightest upper-bound on $r$ that the WMAP team
quote is $r \leq 0.24$ ($2 \sigma$). This constraint is derived from a
combination of the WMAP data with both large scale structure
measurements and the HST key project constraint on the Hubble
constant. It is clear that measurements of the $TT$ and $TE$ CMB
spectra in combination with other astrophysical probes currently play
a significant role in constraining the value of $r$. However, we note
that the constraints obtained from $B$-mode polarization alone are
already comparable to the combined constraints from all other
cosmological probes and are likely to overtake them with the next
generation of CMB polarization experiments.

In Table \ref{tab2} we quote the goodness-of-fits for the various
analyses and we quote the weights for the hyper-parameter analysis
in Table \ref{tab3}. In Table \ref{tab2}, $E(n)$ is the
expectation value for each fit, calculated using Eqs.
(\ref{chi2-expectation}) and (\ref{expectation-hyper}). If the
model provides a good fit to the data, the minimum $\chi^{2}$ over
the expectation value should be well within the range
($1-\sqrt{V(n)}/E(n)$, $1+\sqrt{V(n)}/E(n)$). Examining the table,
we see that the SFI model can fit the BICEP data well, but that
the fit to the QUaD data is relatively poor. This poor fit to the
QUaD data adds further weight to our conclusion above regarding
the anomalous power in the QUaD results in the range $300 < l <
500$.

\begin{table*}[tbp]
\begin{centering}
\begin{tabular}{|c|c|c|c|c|c|}\hline
 Models & Sampling & \multicolumn{3}{|c|}{Conventional $\chi^{2}$} &
 hyper-parameter $\chi^{2}$
\\ \cline{3-5} & Range & BICEP & QUaD (tot) & BICEP+QUaD &
BICEP+QUaD
\\\cline{1-6} SFI: $r$ & $(-1.0,10.0)$ & $
0.01^{+0.31+0.68}_{-0.26-0.49}$ & $10.0^{\times}_{-3.0-9.70}$ &
$0.03^{+0.32+0.70}_{-0.27-0.50}$ &
$0.02^{+0.31+0.75}_{-0.26-0.51}$
\\\cline{1-6} LQG: $m$ [$10^{-8}$M$_{\rm pl}$]&
$(0.01,10^{2})$ & $0.18^{+1.16+5.74}_{\times}$ &
$47.5^{\times}_{-27.7 \text{  }\times}$ &
$0.22^{+1.14+5.72}_{\times}$ & $0.20^{+1.16+5.96}_{\times}$
\\\cline{2-6} $k_{*}$ [$10^{-4}$Mpc$^{-1}$] & $(0.1,10^{2})$ & $1.07^{+1.36+5.88}_{\times}$
& $ 53.4^{+131.63 \text{  } \times}_{-23.3 \text{  } \times}$ &
$1.07^{+1.37+5.89}_{\times}$ & $1.07^{+1.36+5.98}_{\times}$
\\\cline{1-6} Cosmic String & $(10^{-3},10^{2})$ & $(0.001)^{+5.586+9.961}_{\times}$ & tot: $7.60^{+1.38+2.63}_{-1.56-3.60}$
& & \\ \cline{4-6} $G \mu \times 10^{7}$ & & & (l$>$500):
$4.32^{+2.41+4.25}_{\times}$ & $2.98^{+2.82+4.74}_{\times}$&
$2.32^{+3.45+5.69}_{\times}$
\\\hline
\end{tabular}%
\caption{The best-fit values for the parameters, and the $1\sigma$ and
  $2\sigma$ CL for the scalar field inflation (SFI), Loop
  Quantum Gravity (LQG) and cosmic string models. For the SFI and LQG
  models, we use all the QUaD data and combine these with BICEP using
  both a conventional joint $\chi^{2}$ and the hyper-parameter
  approach ($\chi^2_{\text{hyper}}$). For the cosmic strings model,
  in addition to the entire QUaD data set, we also examine the effect
  of removing the $l < 500$ QUaD data points. When combining the QUaD
  and BICEP data for the cosmic strings model, we also restrict the
  QUaD data to $l > 500$ (see the discussion in section
  \ref{constraints-string}). The notation ``$\times$" indicates that the
  values of the parameters are out of sampling range.} \label{tab1}
\end{centering}
\end{table*}

\begin{table*}[tbp]
\begin{centering}
\begin{tabular}{|c|c|c|c|c|}\hline
 Goodness &  \multicolumn{3}{|c|}{Conventional $\chi^{2}$} &
 hyper-parameter $\chi^{2}$
\\ \cline{2-5} of fits & BICEP & QUaD (tot) & BICEP+QUaD &
BICEP+QUaD
\\\cline{1-5} SFI: $E(n)$ & 8 & 22 & 31 & 82.58
\\\cline{2-5} $\chi^{2}_{min}/E(n)$ & 1.00 & 1.64 & 1.56 & 1.26
\\\cline{2-5} Good-fits range & (0.5,1.5) & (0.70,1.30) & (0.75,1.25)
& (0.89,1.11)
\\\cline{1-5} LQG: $E(n)$ & 7 & 21 & 30 & 75.49
\\\cline{2-5} $\chi^{2}_{min}/E(n)$ & 1.10 & 1.76 & 1.60 & 1.22
\\\cline{2-5} Good-fits range & (0.47,1.54) & (0.69,1.31) & (0.74,1.26)
& (0.90,1.10)
\\\cline{1-5} CosStr: $E(n)$ & 8  & 18 ($l>$500)  & 27 (QUaD \text{} $l>$500) &
66.6 (QUaD \text{} $l>$500)
\\\cline{2-5} $\chi^{2}_{min}/E(n)$ & 1.0 & 1.35 & 1.21 & 1.08
\\\cline{2-5} Good-fits range & (0.5,1.5) & (0.67,1.33) & (0.73,1.27)
& (0.90,1.11)
\\\hline
\end{tabular}%
\caption{Reduced $\chi^{2}$ as an indication of the goodness-of-fit
  for each analysis. $E(n)$ is the expectation value for each fit.
  If the model provides a good fit to the data, the values of
  $\chi^{2}_{min}/E(n)$ should be within the range
  ($1-\sqrt{V(n)}/E(n)$, $1+\sqrt{V(n)}/E(n)$).} \label{tab2}
\end{centering}
\end{table*}

\begin{table*}[tbp]
\begin{centering}
\begin{tabular}{|c|c|c|}\hline
 $\alpha_{\text{eff}}=n_{\text{A}}/\chi^{2}_{\text{A}}$ & BICEP
 & QUaD
\\ \cline{1-3} SFI & 1.13 & 0.64
\\ \cline{1-3} LQG & 1.17 & 0.62
\\ \cline{1-3} CosStr & 1.13 & 0.95
\\\hline
\end{tabular}%
\caption{The value of the effective hyper-parameters
$\alpha_{\text{eff}}=n_{\text{A}}/\chi^{2}_{\text{A}}$ for the BICEP
  and QUaD data which reflect the relative weights assigned to each
  data set.}
\label{tab3}
\end{centering}
\end{table*}

\subsection{Prospects for future observations}
\label{fisher-sfi}

In this section, we discuss the detection capabilities of future CMB
experiments. In order to forecast the error bars of the
parameters $r$ and $n_t$ in the fiducial models, we use the
Fisher matrix technique, introduced in Section \ref{fishermatrix}.
\begin{figure}[tbp]
\centerline{\includegraphics[bb=0 0 471
450,width=3.4in]{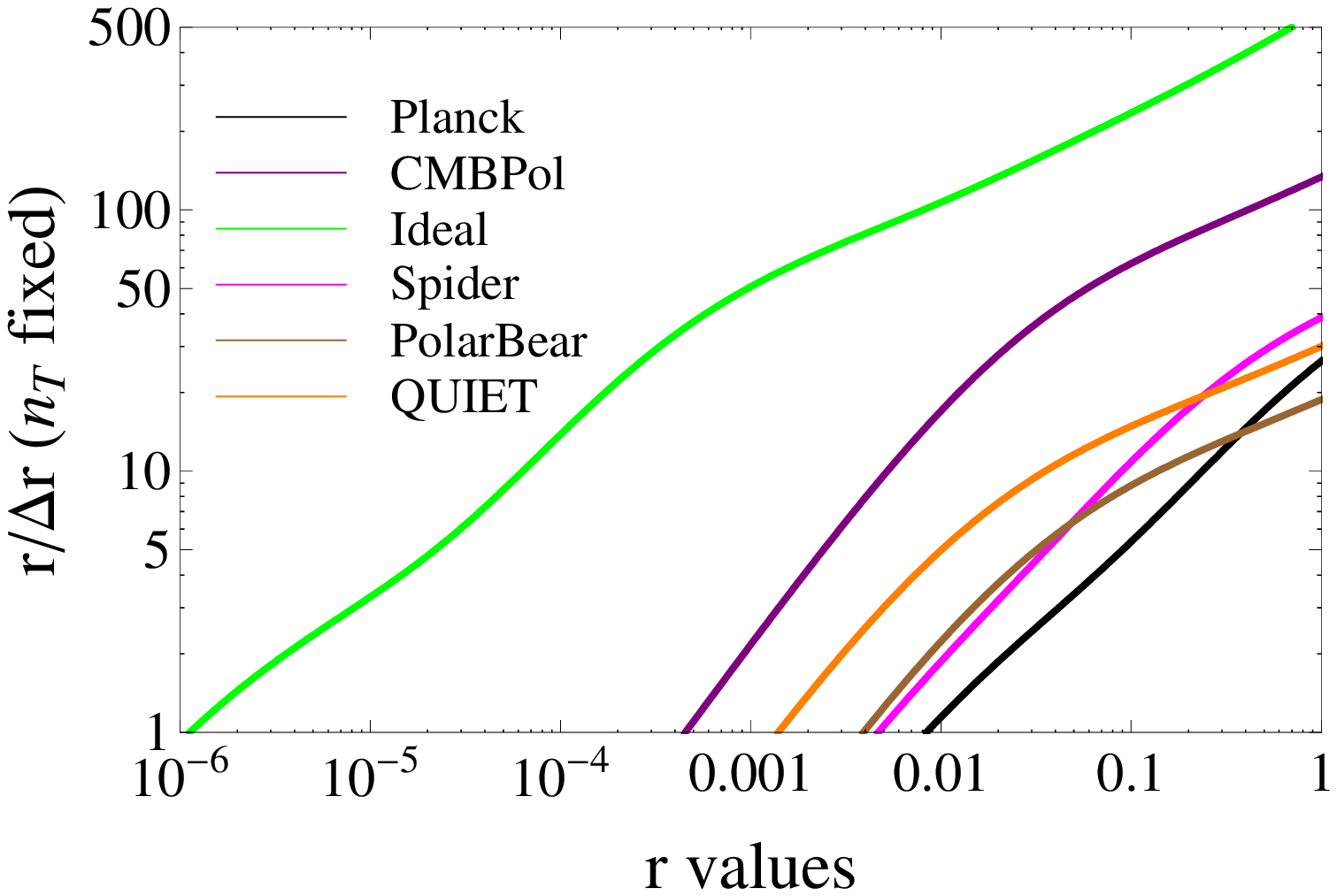}
\includegraphics[bb=0 0 471 450,width=3.4in]{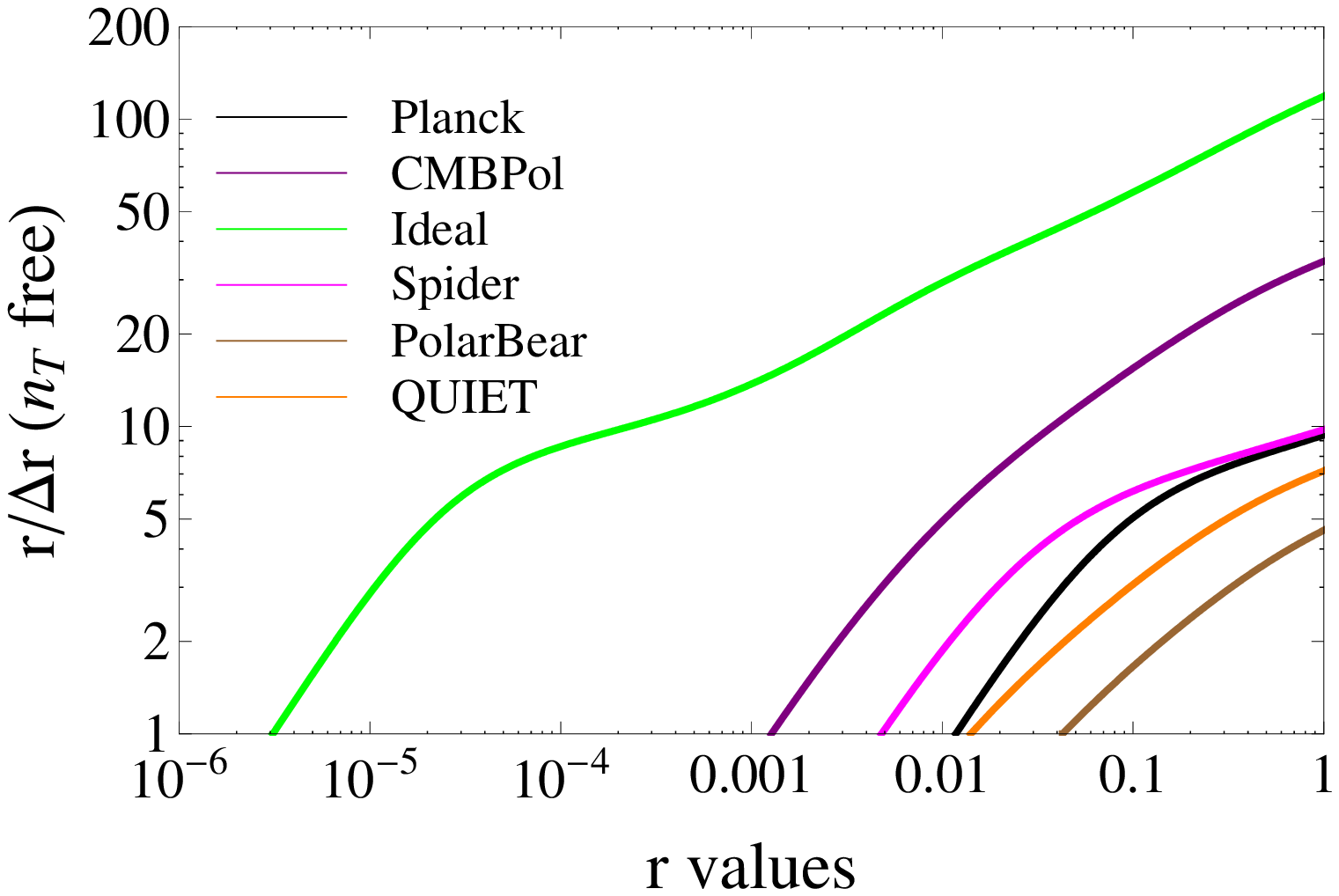}}
\caption{The signal-to-noise ratio $r/ \Delta r$ for different
experiments, calculated using the
Fisher matrix of Eq. (\ref{falphabeta}). Left: The parameter $r$ is
treated as a free parameter but $n_t$ is kept fixed at its fiducial
value; Right: Both $r$ and $n_{t}$ are treated as free parameters.}
\label{deltarnt1}
\end{figure}
\begin{figure}[tbp]
\centerline{\includegraphics[bb=0 0 531
352,width=3.4in]{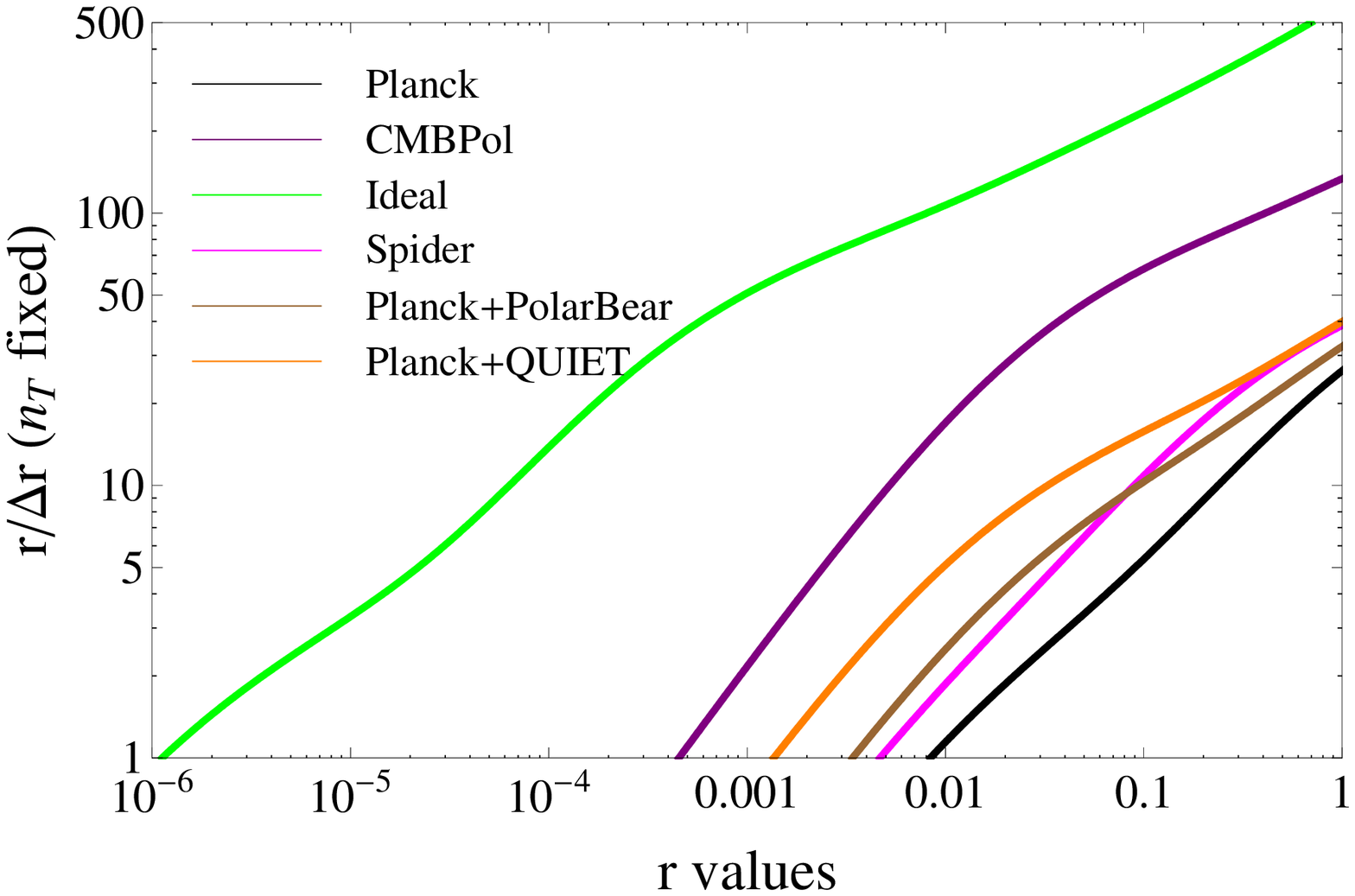}
\includegraphics[bb=0 0 465 309,width=3.4in]{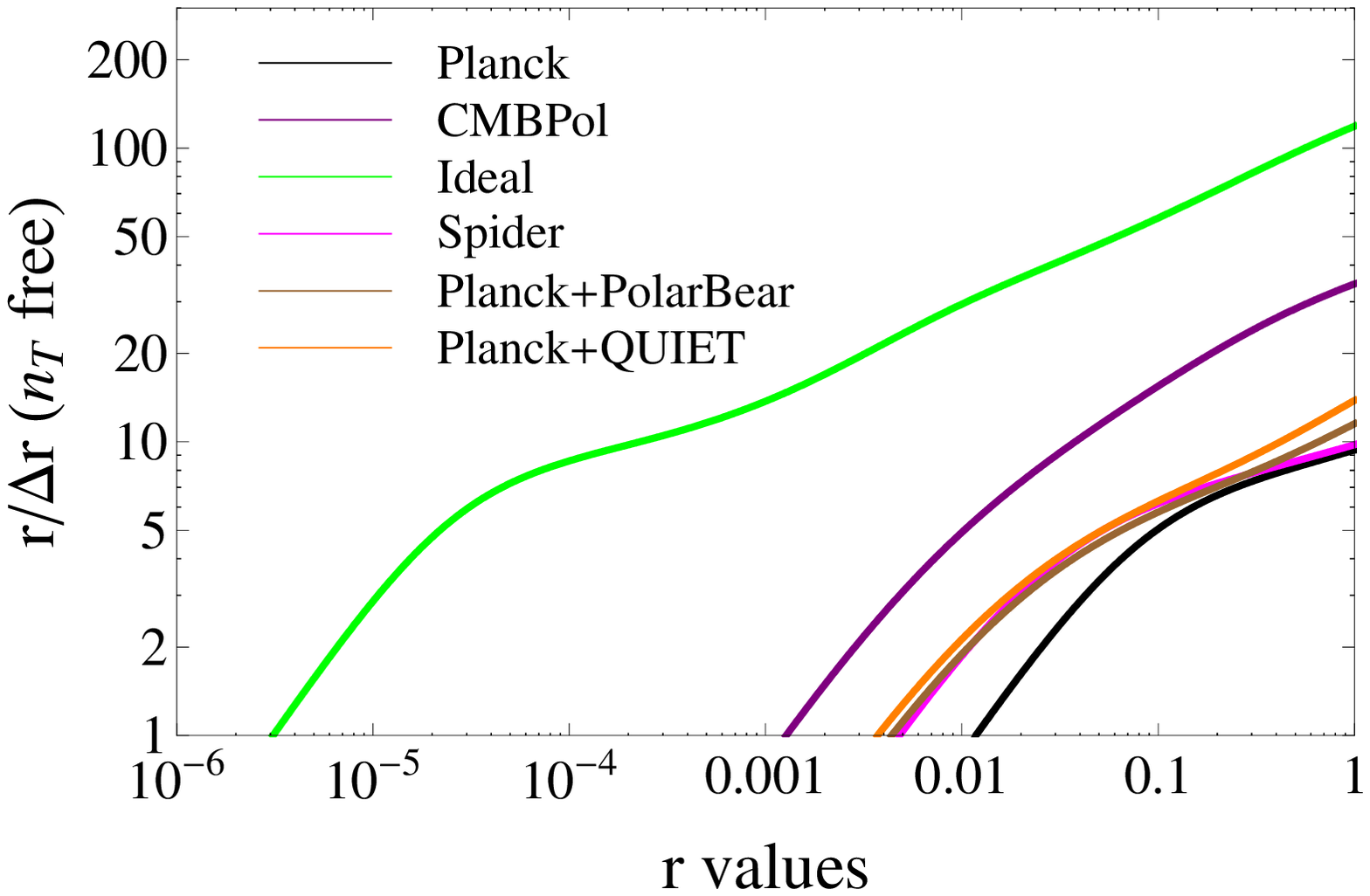}}
\caption{The signal-to-noise ratio $r/ \Delta r$ for different
experiments, and the combination of Planck and ground-based
experiments, calculated using the Fisher matrix of Eq.
(\protect\ref{falphabeta}). Left: The parameter $r$ is
treated as a free parameter but $n_t$ is kept fixed at its fiducial
value; Right: Both $r$ and $n_{t}$ are treated as free parameters.}
\label{deltarnt2}
\end{figure}
\begin{figure}[tbp]
\centerline{\includegraphics[bb=0 0 471
450,width=3.4in]{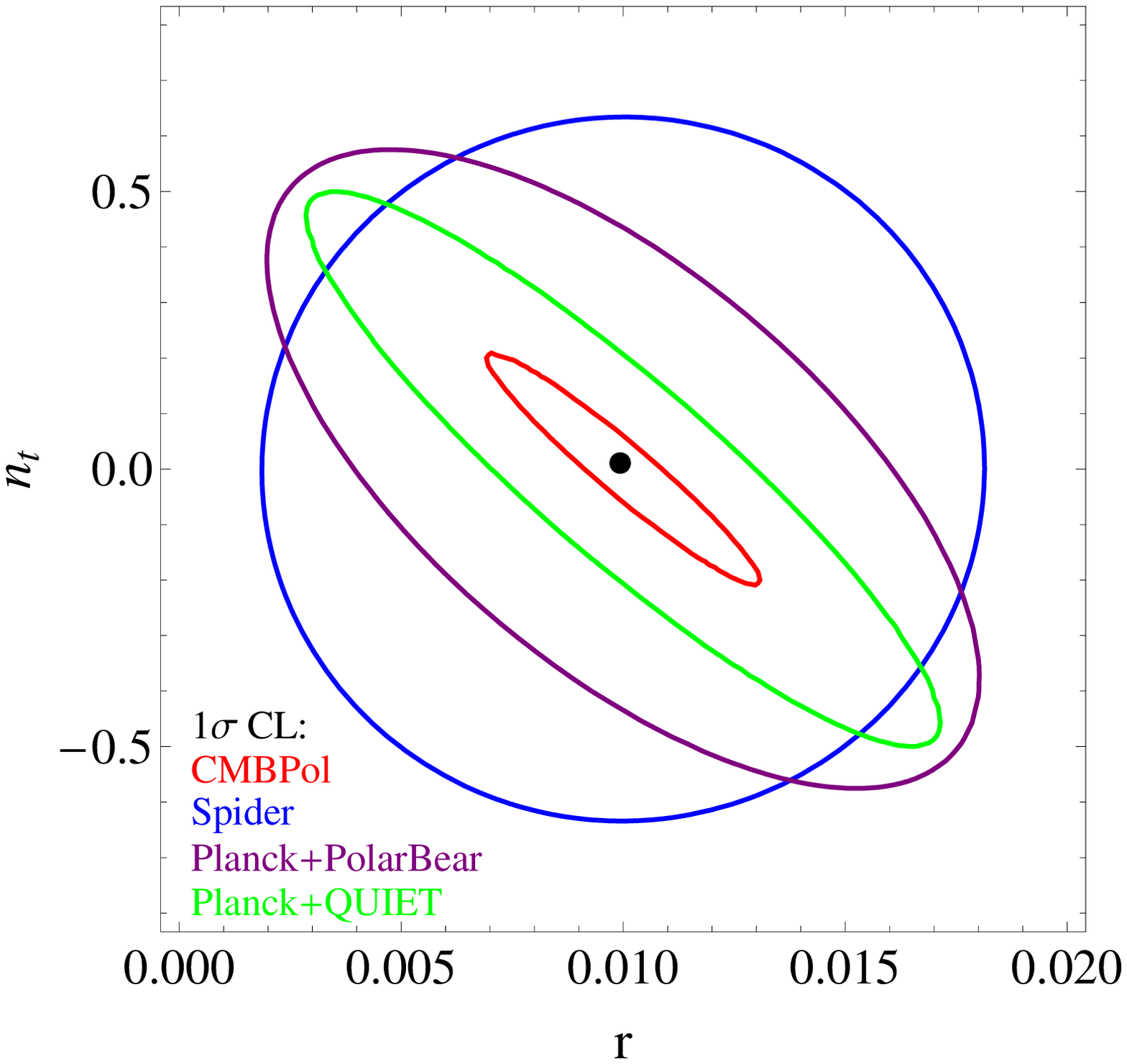}\includegraphics[bb=0 0 495
487,width=3.28in]{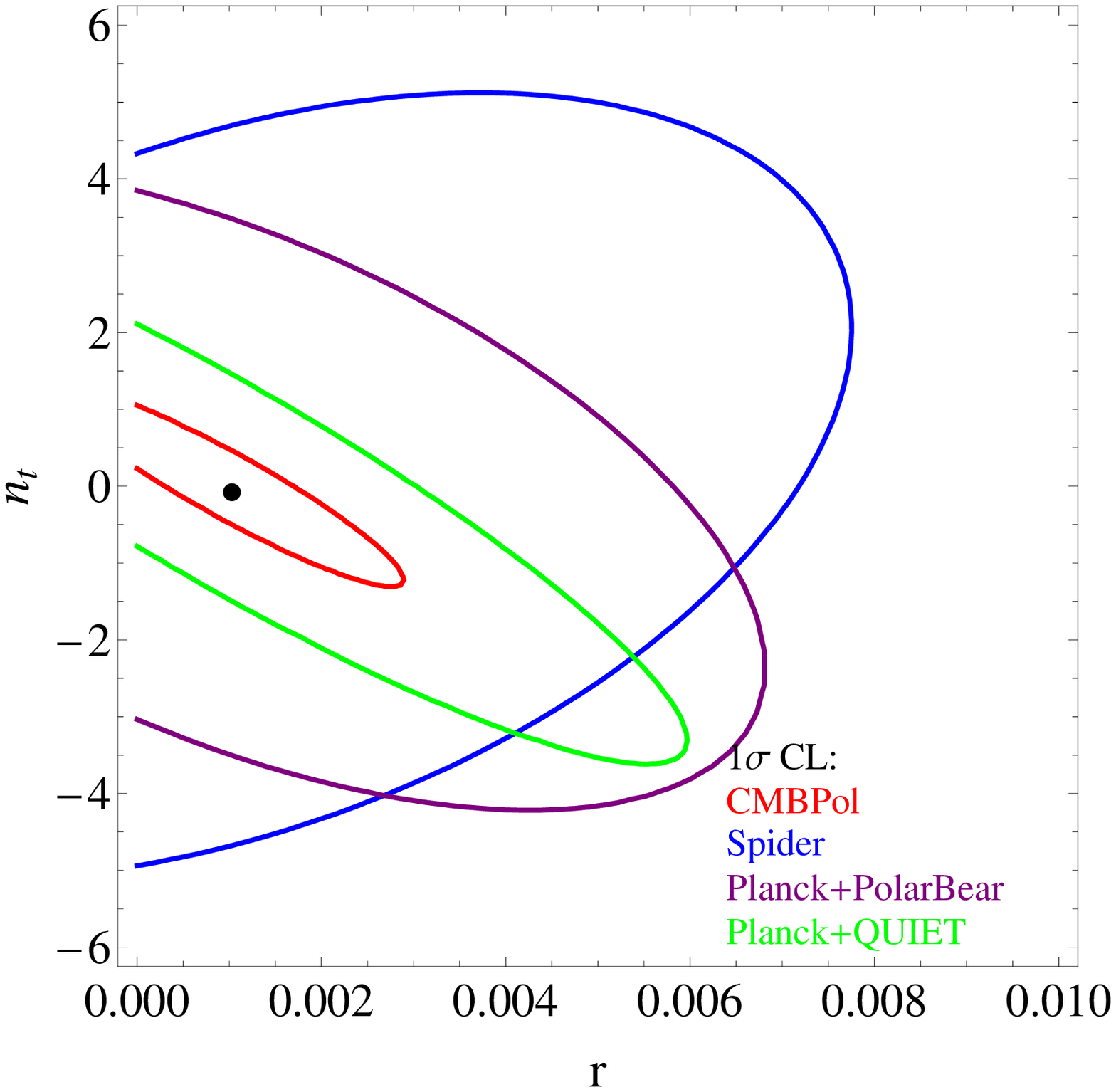}} \caption{Forecasted joint
  constraints on the parameters of the SFI model. The contours
  indicate the 68\% (1$\sigma$) confidence levels. The input models
  are indicated by the black points (Left: $r=0.01$ and $n_{t}=0$, Right:
  $r=0.001$ and $n_{t}=0$).}
\label{sficontours}
\end{figure}
\begin{figure}[tbp]
\centerline{\includegraphics[bb=0 0 548
354,width=3.4in]{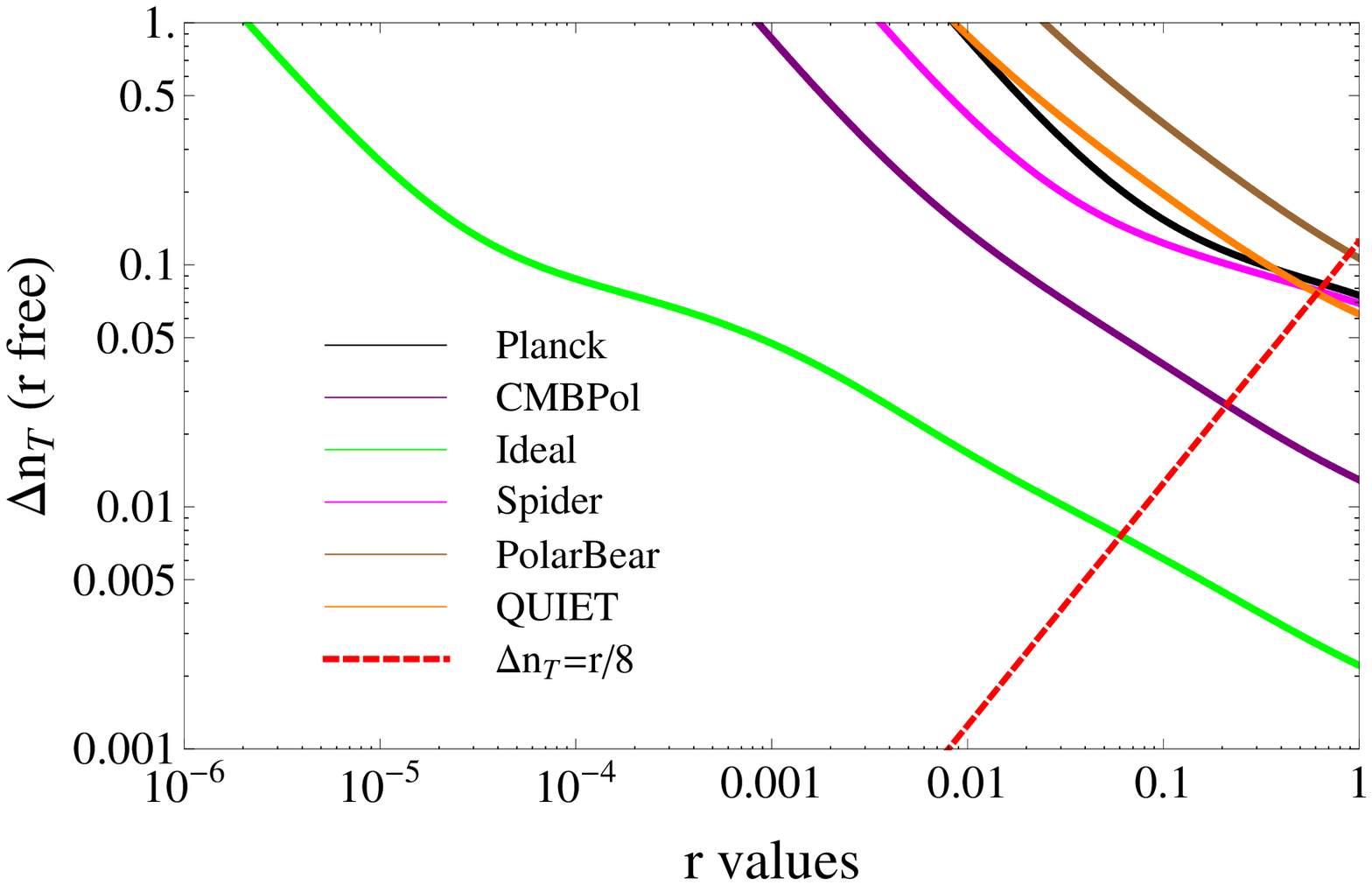}\includegraphics[bb=0 0 547
355,width=3.4in]{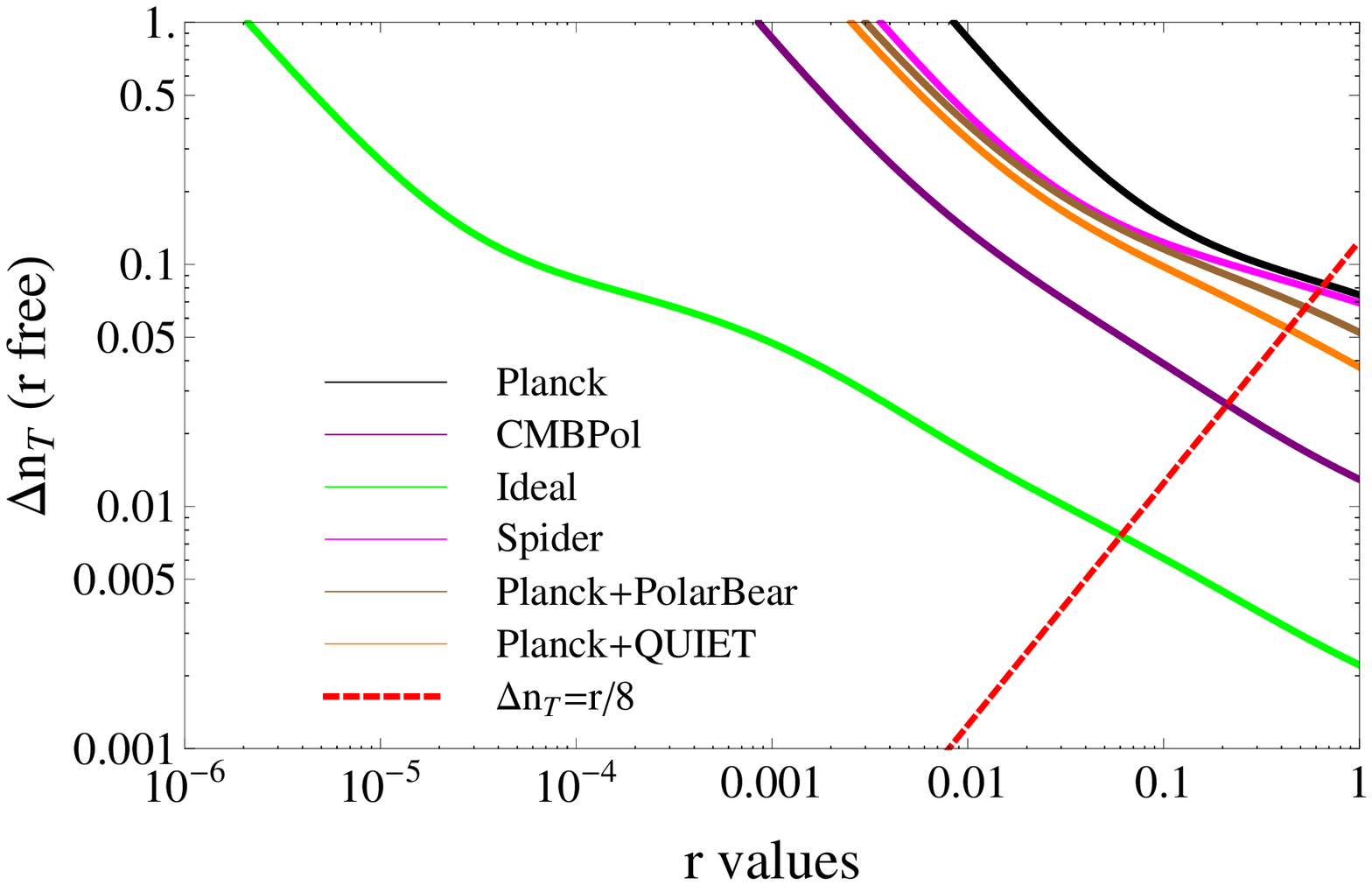}} \caption{Forecasted uncertainty on
  the tensor spectral index, $\Delta n_{t}$, for different experiments,
  calculated using the Fisher matrix of
  Eq. (\protect\ref{falphabeta}). Here, both $r$ and $n_{t}$ are
  treated as free parameters.} \label{deltarnt3}
\end{figure}

In Fig.~\ref{deltarnt1}, we plot the signal-to-noise ratio ($r/\Delta
r$) for a detection of tensors as a function of the fiducial value of
$r$, for a number of current and forthcoming experiments.  In the left
panel, we only consider $r$ as the free parameter, and keep $n_t$
fixed at $n_t=0$. We see that the Planck satellite can potentially
detect the signal of the tensor perturbations at more than $3\sigma$
confidence level if $r>0.05$. For $r=0.1$, the value of $r/\Delta r$
becomes $5$ which would constitute a robust detection. These results
are consistent with those presented in \cite{Efstathiou09,zbg2010}. The
predicted constraints for PolarBear, QUIET and Spider are somewhat
tighter with $r/\Delta r>3$ for models with $r>0.02$. The predicted
constraints for the proposed CMBPol mission suggest that tensor
perturbations could be detected (at the 3$\sigma$ level) for values of
$r$ as low as $r \sim 0.002$. Such a measurement would provide an
excellent opportunity to differentiate between various inflationary
models. We also find that for the ideal CMB experiment which includes
only a residual lensing noise contribution (after de-lensing), the SFI
primordial signal could be detected (at $> 3\sigma$) only if
$r>10^{-5}$ is satisfied.

In Fig.~\ref{deltarnt2}, we also plot the signal-to-noise ratio for
the combination of Planck with the ground-based experiments (PolarBear
and QUIET). The former is sensitive to the $B$-mode signal at the lowest
multipoles $\ell<20$, while the latter are sensitive to the recombination
peak of $C_{l}^{BB}$ at $l\sim 80$. Similar to \cite{zb}, we find that
the combination of these experiments yields little formal improvement in the
signal-to-noise of the detection compared with the capabilities of the
ground-based experiments on their own. However, a detection of
\emph{both} the recombination bump (e.g. from ground-based
experiments) \emph{and} the reionization bump (e.g. from Planck) would
constitute much more compelling evidence for tensors than either
detection would constitute on its own.

In the right hand panels of Fig.~\ref{deltarnt1} and Fig.~\ref{deltarnt2},
we have plotted the results for the case where we treat both $r$ and
$n_t$ as free parameters. Comparing with the corresponding results in
the left panels, we find (in agreement with previous works,
e.g. \cite{cmbpol, nt2}) that the signal-to-noise ratios become much
smaller due to correlations between the parameters. These correlations
were investigated in some detail by \cite{zb} who also explored the
optimal choice of pivot scale for which the two parameters become
decorrelated.

In Fig.~\ref{deltarnt3}, we plot the values of $\Delta n_t$ as a
function of $r$ for various cases. Here we find that, for Planck,
PolarBear, QUIET and Spider, the constraint on $n_t$ is relatively
weak unless the value of $r$ is very large. For example, the predicted
constraint for Planck is $\Delta n_t=0.13$, and for QUIET is $\Delta
n_t=0.18$ for a model with $r=0.1$. The combination of Planck
and QUIET could in principle do somewhat better with $\Delta
n_t=0.08$. The proposed CMBPol mission could achieve $\Delta
n_t=0.04$ while the limiting value of the ideal experiment is $\Delta
n_t=0.007$, which is comparable to the current constraint on the
scalar spectral index $n_s$ \cite{Komatsu10}. For lower values of $r$,
the predicted constraints are correspondingly weaker. For example, for
the CMBPol mission, the value of $\Delta n_t$ increases from $0.04$ to
$0.1$ if we replace the $r=0.1$ model with $r=0.01$.

Testing the consistency relation $n_t=-r/8$ (see Eq.
(\ref{consistent-relation})) is potentially one of the most
powerful ways to test the general SFI scenario. To assess whether
future experiments might achieve this goal, in
Fig.~\ref{deltarnt3} we compare the values of $\Delta n_t$ with
$r/8$. If $\Delta n_t<r/8$, then the constraint on $n_t$ is tight
enough to allow the consistency relation to be tested. We find
that $\Delta n_t<r/8$ is satisfied only if $r>0.23$ for the CMBPol
experiment, and only if $r>0.06$ for the ideal experiment. An
observational confirmation of the consistency relation is
therefore extremely unlikely to be achieved with any of the
currently envisaged future experiments.

\subsection{Single-field slow-roll inflationary models}
\label{sfi-model-discussion} In this subsection, we discuss four
types of single field slow-roll inflationary models, categorized
by \cite{cata}, and the implications for these models of the results
presented in the previous section.

In single field slow-roll inflationary models, the observables
depend on three slow-roll parameters \cite{slow-roll}
\begin{equation}  \label{p}
\epsilon_V\equiv\frac{{\rm M}_{\mathrm{pl}}^2}{2}\left(\frac{V^{\prime }}{V}%
\right)^2~, ~~~~~~ \eta_V\equiv
{\rm M}_{\mathrm{pl}}^2\left(\frac{V^{\prime
\prime }}{V}\right)~, ~~~~~~ \xi_V\equiv {\rm M}_{\mathrm{pl}}^4\left(\frac{%
V^{\prime }V^{\prime \prime \prime }}{V^2}\right)~,
\end{equation}
where $V(\phi)$ is the inflationary potential, and the prime denotes
derivatives with respect to the field $\phi$. Here, $\epsilon_V $
quantifies the ``steepness" of the slope of the potential, $\eta_V$
measures the ``curvature" of the potential, and $\xi_V$ quantifies the
``jerk". Since the potential is fairly flat in the slow-roll inflation
models, these three parameters must be much smaller than unity for
inflation to occur. One of the important predictions of SFI models is
that the scalar perturbations are nearly scale-invariant, which has
already been confirmed by WMAP results \cite{Komatsu10}.

In SFI models, a standard slow-roll analysis yields the following relations
\begin{equation}  \label{relation}
n_t=-\frac{r}{8},~~ n_s=1+2\eta_V-6\epsilon_V, ~~
r=\frac{8}{3}(1-n_s)+\frac{16}{3}\eta_V,~~\alpha_s=-24\epsilon^2_V+16\epsilon_V
\eta_V-2\xi_V, \label{sfi-relation}
\end{equation}
where $n_s$ is the tilt of primordial scalar power spectrum, and
$\alpha_s = d n_s / d \ln k$ is the ``running'' of $n_s$. These
formulae relate the tensor parameters $n_t$ and $r$ to the scalar
parameters $n_s$ and $\alpha_s$; the latter can be constrained through
CMB and large scale structure observations. As shown in
Eq. (\ref{relation}), the relation between $r$ and $n_s$ involves the
slow-roll parameter $\eta_V$ which in turn depends on the specific
inflationary potential.

The strength of the primordial tensor perturbations depends on the
value of $r$. Observations have yielded quite tight constraints on
$n_s$, but we currently only have upper limits on the value of
$r$. The relation between $n_s$ and $r$ depends on the specific
inflationary model, and different models predict very different values
for $r$. In the following discussion, we categorize SFI models into
four classes based on different regimes for the curvature of the
potential $V(\phi)$, and discuss their individual constraints.

\emph{Case A: negative curvature models $\eta_V<0$}

The negative $\eta_V$ models arise from a potential of spontaneous
symmetry breaking. One type of often-discussed potentials is the form $%
V=\Lambda^4\left[1-(\phi/\mu)^p\right]$ ($p\geq2$). This type of
model predicts a red tilt in the scalar spectrum ($n_s<1$), which is
consistent with the WMAP 7-year results \cite{Komatsu10}. In addition,
these models predict relatively small values for $r$. For the model
with $p=2$ in Ref. \cite{cata},
\begin{equation}  \label{casea}
r\simeq 8(1-n_s)e^{-N_{e}(1-n_s)}~,
\end{equation}
where $N_{e}$ is the number of e-folds, taken to be in the range
$N_{e}\in[40,~70]$ based on current observations of the
CMB~\cite{N,Komatsu10}. Here we choose the value $N_{e}=60$. Using
the result $n_s=0.963\pm0.012$ \cite{Komatsu10} yields the
constraint $r\in[0.021,~0.045]$. From Fig.~\ref{deltarnt1} (right
panel), we see that this is close to or even beyond the
sensitivity range of the Planck satellite, but is within the
sensitivity ranges of PolarBear, QUIET, Spider and CMBPol. In
other models with $p>2$, the predicted values of $r$ are much
smaller than that of the model with $p=2$.

\emph{Case B: small positive curvature models $0\leq\eta_V\leq2\epsilon_V$}

Two example potentials in this case are the monomial potentials $%
V=\Lambda^4(\phi/\mu)^p$ with $p\geq2$ for $0<\eta_V<2\epsilon_V$ and the
exponential potential $V=\Lambda^4\exp(\phi/\mu)$ for $\eta_V=2\epsilon_V$.
In these models, to first order in slow roll, the scalar index is always
red $n_s<1$ and the following constraint on $r$ is satisfied
\begin{equation}  \label{caseb}
\frac{8}{3}(1-n_s)\leq r\leq8(1-n_s)~.
\end{equation}
Using the result $n_s=0.963$ \cite{Komatsu10} , one finds that $r\in[0.1,~0.3]$,
which is within the sensitivity range of the Planck satellite, as well
as that of forthcoming CMB experiments. Thus, the Planck results may
provide some constraints on these type of models.

\emph{Case C: intermediate positive curvature models $2\epsilon_V<\eta_V%
\leq3\epsilon_V$}

The supergravity-motivated hybrid models have a potential of the form $%
V\simeq\Lambda^4\left[1+\alpha\ln(\phi/Q)+\lambda(\phi/\mu)^4\right]$, up to
one-loop correction during inflation. In this case,
\begin{equation}  \label{cased}
n_s<1~,~~~~~~r>8(1-n_s)~,
\end{equation}
are satisfied. Using the result $n_s=0.963$ \cite{Komatsu10}, one finds
that $r>0.3$, which is slightly in conflict with the current upper
limit $r<0.24$ (WMAP+BAO+$H_{0}$, $2\sigma$ CL) \cite{Komatsu10}.
Fig.~\ref{deltarnt1} shows that this model is also in the sensitivity
range of the Planck satellite.

\emph{Case D: large positive curvature models $\eta_V>3\epsilon_V$}

This class of models has a typical monomial potential similar to
those of Case A, but with a plus sign for the term $(\phi/\mu)^p$:
$V=\Lambda^4\left[1+(\phi/\mu)^p\right]$. This enables inflation
to occur for small values of $\phi<\text{M}_{\text{pl}}$. The
model predicts a blue tilt in the scalar power spectrum $n_s>1$
(Eq. (\ref{sfi-relation})), which is in conflict with current
constraints on $n_s$ unless a running of $n_s$ is allowed
\cite{Komatsu10}. When a running in $n_s$ is included, the WMAP
7-year results suggest a blue power spectrum ($n_s= 1.008 \pm
0.042 $ for $1 \sigma$ CL). Therefore, even though this model is
not favoured by the WMAP 7-year results for the case of constant
$n_s$, it is not excluded when a running of the spectral index is
included. Planck and future CMB experiments should constrain both
$n_{s}$ and $\alpha_s$ to high precision and so should be able to
definitively rule out this model.

\section{Loop Quantum Gravity and its observational probes}
\label{lqg-model}
\subsection{Primordial tensor perturbations in LQG models}
\label{perturbation-loop}

Loop Quantum Gravity is a promising framework for constructing a
quantum theory of gravity in theoretical physics. Based on the
reformulation of General Relativity as a kind of gauge theory obtained
by \cite{Sen81,Ashtekar96}, LQG is now a language and a dynamical
framework which leads to a mathematically coherent description of the
physics of quantum spacetime \cite{Grain09}.  Constraining LQG
theories experimentally is challenging because the quantum geometrical
effect can only be tested at very high energy scales, beyond the reach
of current accelerator experiments. In this section, we will calculate
the possible observational signature of LQG in the CMB sky, which
opens a new window for cosmological tests of quantum gravity.


There are two main quantum corrections in the Hamiltonian of LQG when
dealing with the semi-classical approach, namely holonomy corrections
and ``inverse volume" corrections \cite{Grain09,Mielczarek10}. The
holonomy corrections lead to a dramatic modification of the Friedmann
equation as \cite{Singh06}
\begin{equation}
H^{2}=\frac{8 \pi G}{3} \rho \left( 1-\frac{\rho}{\rho_c} \right),
\label{loop-hubble}
\end{equation}
where $\rho$ is the energy density, and $\rho_c$ is the critical
energy density,
\begin{equation}
\rho_c=\frac{4 \sqrt{3}}{\gamma^3} \text{M}_{\text{pl}}^4 \simeq
507.49 \, \text{M}_{\text{pl}}^4. \label{rhoc}
\end{equation}
Here $\gamma=0.239$ is the Barbero-Immirzi parameter, which is derived
from the computation of the black hole entropy \cite{Meissner04}. Note
that we use the reduced Planck mass in our calculation.

A generic picture for this model with the holonomy correction is the
bouncing behavior exhibited when the energy density of the Universe
approaches $\rho_c$. The negative sign in Eq.~(\ref{loop-hubble}) is
an appealing feature in the framework of LQG such that the repulsive
quantum geometry effect becomes dominant in the Planck region
\cite{Grain09,Mielczarek10}. This triggers a contraction period before
the bounce, during which time the Hubble parameter is negative and the
Hubble radius is shrinking. As a result, the perturbation modes on the
largest scales crossed the Hubble horizon and froze out during the
contracting period, until the end of the contracting stage when the
Hubble horizon increased again. For the very large scale modes, this
pre-inflationary bounce may imprint distinctive features in the CMB
sky, since they stretched out of the horizon at very early times
\cite{Grain09,Mielczarek10}.

Unfortunately, the power spectrum for the scalar perturbations is
somewhat hard to obtain because in the case of the holonomy
correction, the anomaly free equations are still to be found
\cite{Mielczarek10b}. Therefore, in the following discussion, we will
focus on the tensor power spectrum of LQG and pursue the constraints
obtainable on the model from $B$-mode observations.

Due to the pre-inflationary contracting period, the tensor power
spectrum for LQG can be calculated numerically. In
\cite{Mielczarek10}, a simple parameterized form of the power
spectrum is introduced as follows
\begin{equation}
P_{t}=\frac{2}{\pi ^{2}}\left( \frac{H}{\text{M}_{\text{pl}}}\right) ^{2}\frac{1}{%
1+(k_{\ast }/k)^{2}}\left[ 1+\frac{4R-2}{1+(k/k_{\ast })^{2}}\right] ,
\label{pt11}
\end{equation}%
where $H$ is the Hubble parameter during the inflationary stage,
$k_{\ast}$ is the position of the highest peak in the power
spectrum, and the quantity $R$ is related to the mass of the
scalar field as
\begin{equation}
R=(8\pi )^{0.32}\left[ \frac{\text{M}_{\text{pl}}}{m}\right] ^{0.64}.
\end{equation}%
It is interesting to note that Eq. (\ref{pt11}) reduces
to the SFI result of Eq. (\ref{pt1}) for $k_{\ast}\rightarrow 0$.
In this paper, for simplicity, we consider the tensor power
spectrum (Eq. \ref{pt11}) with a constant $H$ in the early stage of
inflation, which corresponds to the specific case of de Sitter
inflation.

In this model, we assume that inflation is driven by the
potential $V(\phi)=m^2 \phi^2/2$. The Hubble parameter is related
to this potential via
\begin{equation}
H^{2}=\frac{1}{3{\rm M}_{\rm pl}^{2}}V(\phi )=\frac{1}{6{\rm M}_{\rm pl}^{2}}m^{2}\phi
_{i}^{2}, \label{Hubble}
\end{equation}%
where $\phi _{i}$ is the initial value of the scalar field at the
beginning of inflation. The number of e-folds can be calculated as
(using the slow-roll approximation $3H\dot{\phi}=-V^{\prime }$)
\begin{eqnarray}
N_{e} &=&\int_{i}^{f}Hdt  \nonumber \\
&=&-\frac{1}{{\rm M}_{\rm pl}^{2}}\int_{i}^{f}\frac{V}{V^{\prime }}d\phi   \nonumber \\
&=&-\frac{1}{4{\rm M}_{\rm pl}^{2}}\left( \phi _{f}^{2}-\phi
_{i}^{2}\right).
\end{eqnarray}%
Since $\phi _{i}\gg \phi _{f}$, the above equation is approximately
given by $\phi _{i}^{2} \simeq 4N_{e}{\rm M}_{\rm pl}^{2}.$ The Hubble
parameter is then (from Eq. (\ref{Hubble}))
\begin{equation}
H^{2} \simeq \frac{2}{3}N_{e}m^{2}. \label{hubble-efolds}
\end{equation}
Substituting this into Eq. (\ref{pt11}), we arrive at the
following expression for the tensor power spectrum:
\begin{equation}
P_{t}=\frac{4N_{e}}{3\pi }\left( \frac{m}{\text{M}_{\text{pl}}}\right) ^{2}\frac{1}{%
1+(k_{\ast }/k)^{2}}\left[ 1+\frac{4\times (8\pi )^{0.32}\times
(m/\text{M}_{\text{pl}})^{-0.64}-2}{1+(k/k_{\ast })^{2}}\right] ,  \label{pt2}
\end{equation}%
where $N_{e}$ is the number of e-folds which we fix at $N_{e}\simeq
60$. Finally, we use Eq.~(\ref{clbb-eq}) to project the perturbation
modes onto the CMB sphere to find $C_{l}^{BB}$ for the LQG model.

\begin{figure}[tbp]
\centerline{\includegraphics[bb=0 0 704 450,width=3.4in]{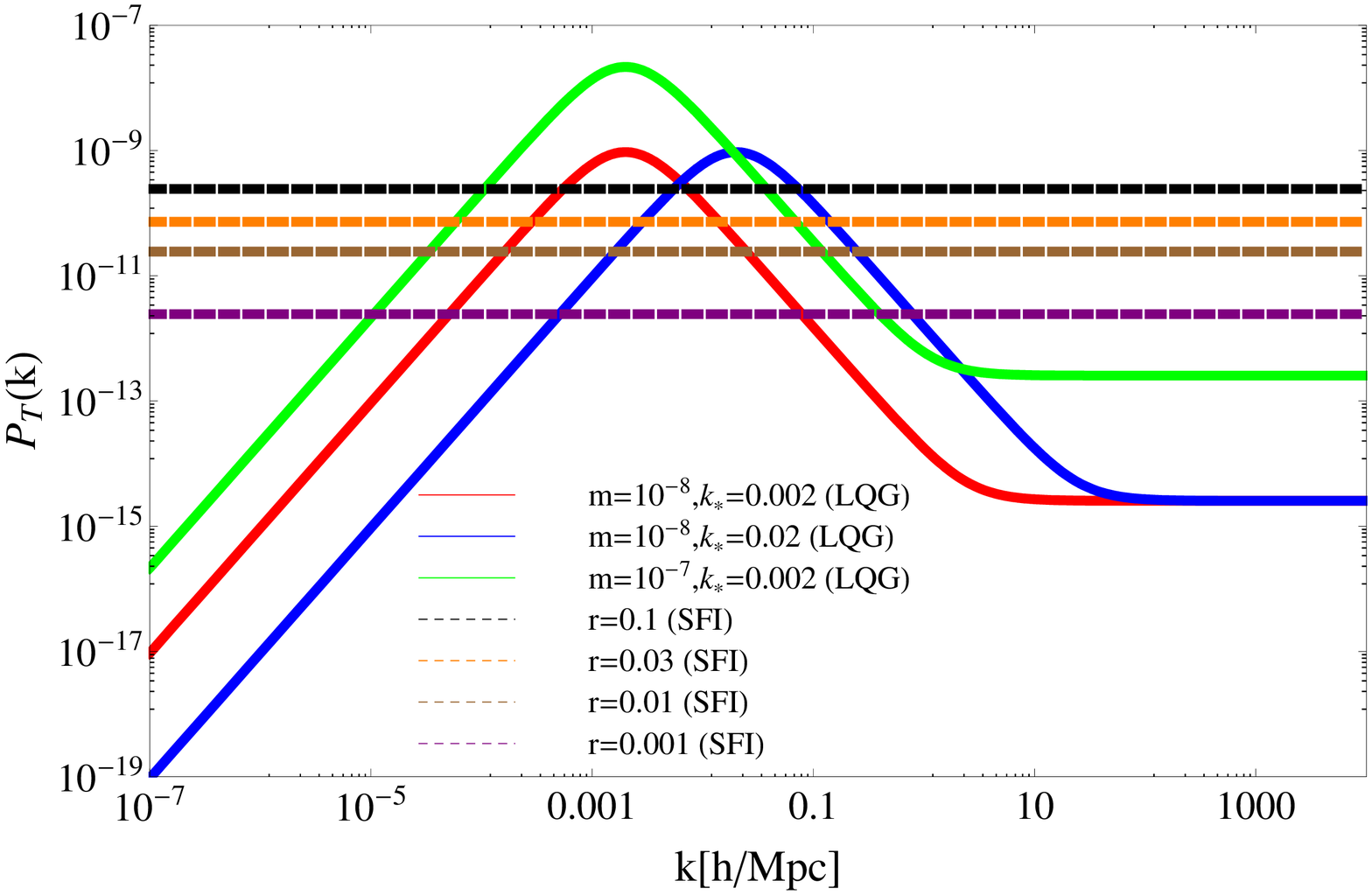}
\includegraphics[bb=0 0 586 378,width=3.4in]{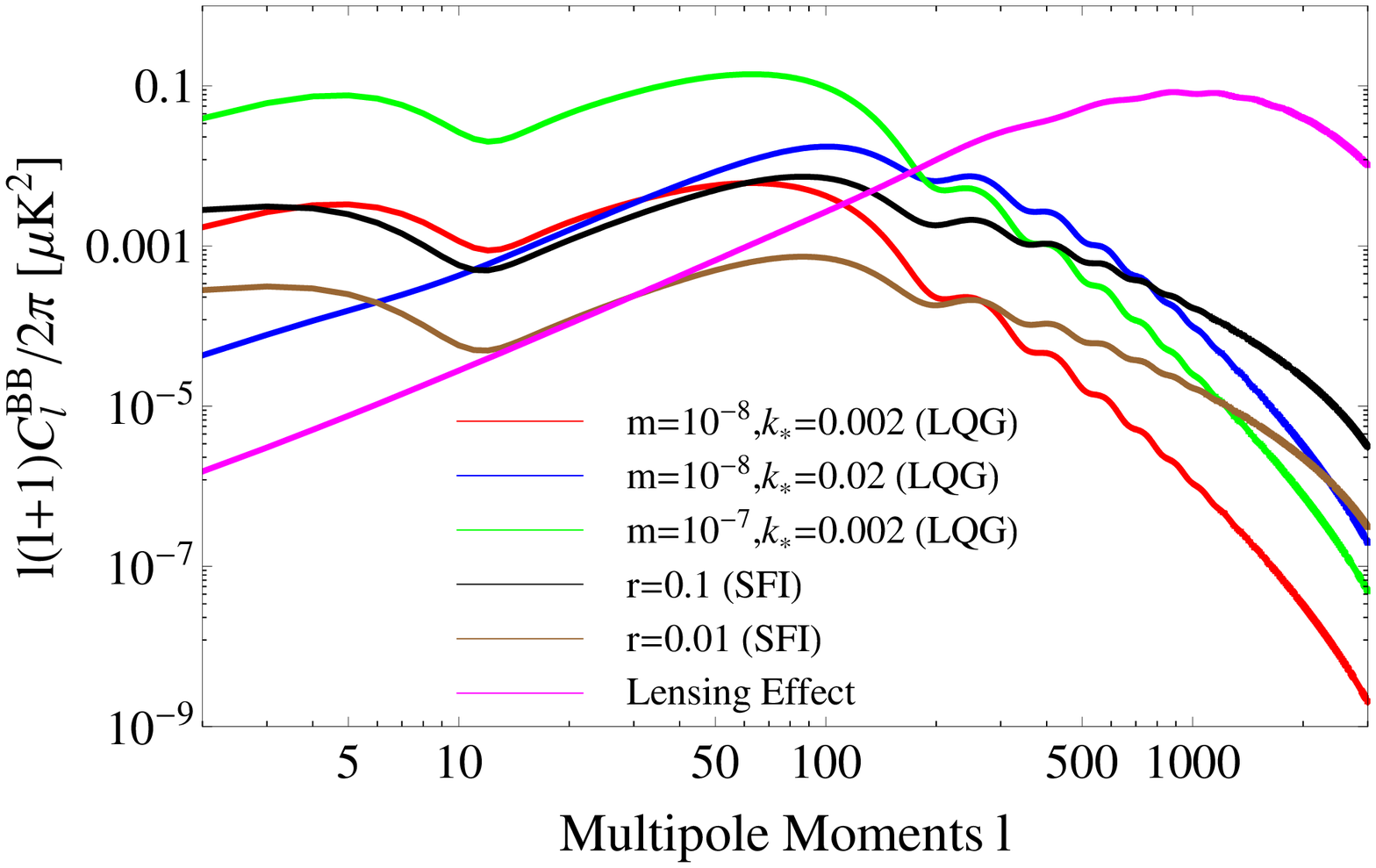}}
\caption{Left: the primordial tensor power spectrum for different
  models of inflation. Right: the corresponding $B$-mode angular power
  spectrum generated by different inflationary models. Note the
  lensing $B$-mode signal which acts as an effective noise for
  detecting the primordial $C_{l}^{BB} $ signal. The units for $m$ and
  $k_{*}$ are $\text{M}_{\text{pl}}$ and $\text{Mpc}^{-1}$
  respectively.} \label{powerspectrum}
\end{figure}

In the left panel of Fig. \ref{powerspectrum}, for a number of
representative sets of parameters, we plot the primordial tensor power
spectrum for the LQG model alongside the signal expected in an SFI
model for a number of different values of $r$. For the SFI model,
since the power spectrum tilt $n_t$ is very small ($n_t=0$ for de
Sitter inflation), the power spectrum is very flat on all scales. In
comparison, the tensor power spectrum of LQG exhibits a bump feature,
which is in fact the signature of the pre-inflationary contraction
period. Very large scale modes stretched out of the Hubble horizon
during the contracting period before inflation, and can be described
by the solution in the Minkowski vacuum $f_k=e^{-ik \eta}/\sqrt{2k}$
\cite{Grain09,Mielczarek10}. Thus, the power spectrum at very large
scale takes the form $P_t(k)\sim k^{3} \left\vert f_k \right\vert ^{2}
\sim k^2$ \cite{Grain09,Mielczarek10}. In contrast, the small scale
modes are well within the Hubble horizon and so the power on small
scales is similar to the scale-invariant power spectrum of the SFI
model. The bump in the power spectrum on larger scales is
characterized by the magnitude of $k_{*}$. As $k_{*}$ increases, the
bump is shifted to smaller scales and vice versa. The amplitude of the
spectrum and the width of the bump are determined by the mass
parameter $m$. We will link these two important parameters to the
energy scale of inflation and the current Hubble horizon scale in the
next subsection.

The right panel of Fig. \ref{powerspectrum} shows that the bump in
the primordial power spectrum results in a peak in the CMB $B$-mode
power spectrum, which is slightly different to that of the SFI
model. In addition, if the peak of the LQG spectrum is normalized
to the same magnitude as that of SFI, the small scale power will
be suppressed in the LQG model, as compared to the SFI model.

\subsection{Constraints from current data}
\label{constraints-loop} In this subsection, we use the BICEP and
QUaD data to constrain the parameters of LQG models. Before we
perform the parameter estimation, we link the two parameters $m$
and $k_{*}$ with the energy scale of inflation, and with the current
Hubble horizon scale.

The parameter ${m}$ relates to the energy scale of inflation in LQG
as follows
\begin{equation}
V^{\frac{1}{4}}=3.02\times 10^{15} \text{GeV} \left(
\frac{m}{10^{-7} \text{M}_{\text{pl}}} \right) ^{{1}/{2}},
\label{energyscaleloop}
\end{equation}
where $10^{15} \, \text{GeV}$ is around the GUT energy scale.
Therefore, a detection of $m>10^{-7} \text{M}_{\text{pl}}$
would strongly suggest that the energy scale of inflation is above the
GUT scale.

The parameter $k_*$ describes the position of the peak in the
primordial power spectrum. We can compare it with the current Hubble
wavenumber, which is $k_H\equiv H_0\simeq 2.33 \times 10^{-4}
\text{Mpc}^{-1}$. If $k_{*} > k_{H}$, then modes with physical
wavelengths ($\lambda_{*}$) equal to the Hubble horizon at the
beginning of inflation will have wavelengths less than the current
Hubble horizon, whereas if $k_* < k_H$ their wavelengths will be
larger than the current horizon scale. Thus, if $k_{*} > k_{H}$, we would
expect to be able to find pre-inflationary fluctuations within our
current Hubble horizon \cite{Grain09,Mielczarek10}. Conversely, as
$k_{*} \rightarrow 0$, the primordial tensor power spectrum
(Eqs. (\ref{pt11}) and (\ref{pt2})) reduces to the scale-invariant
tensor power spectrum as noted above. Therefore, a non-zero detection
of $k_{*}$ would strongly indicate the existence of a bounce and of a
contracting period before inflation.

\begin{figure}[tbp]
\centerline{\includegraphics[bb=0 0 550
364,width=3.4in]{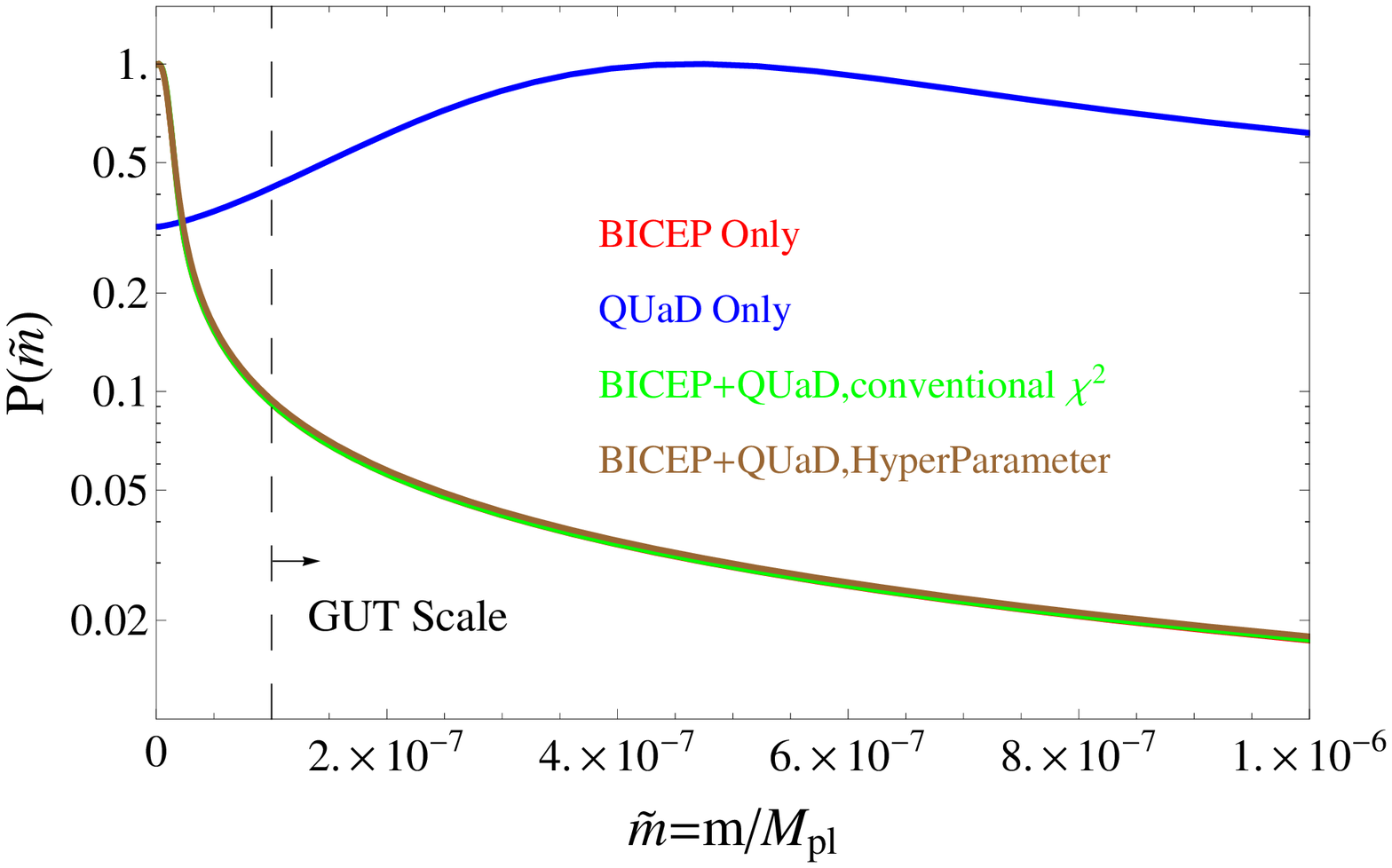}\includegraphics[bb=0 0 755
497,width=3.4in]{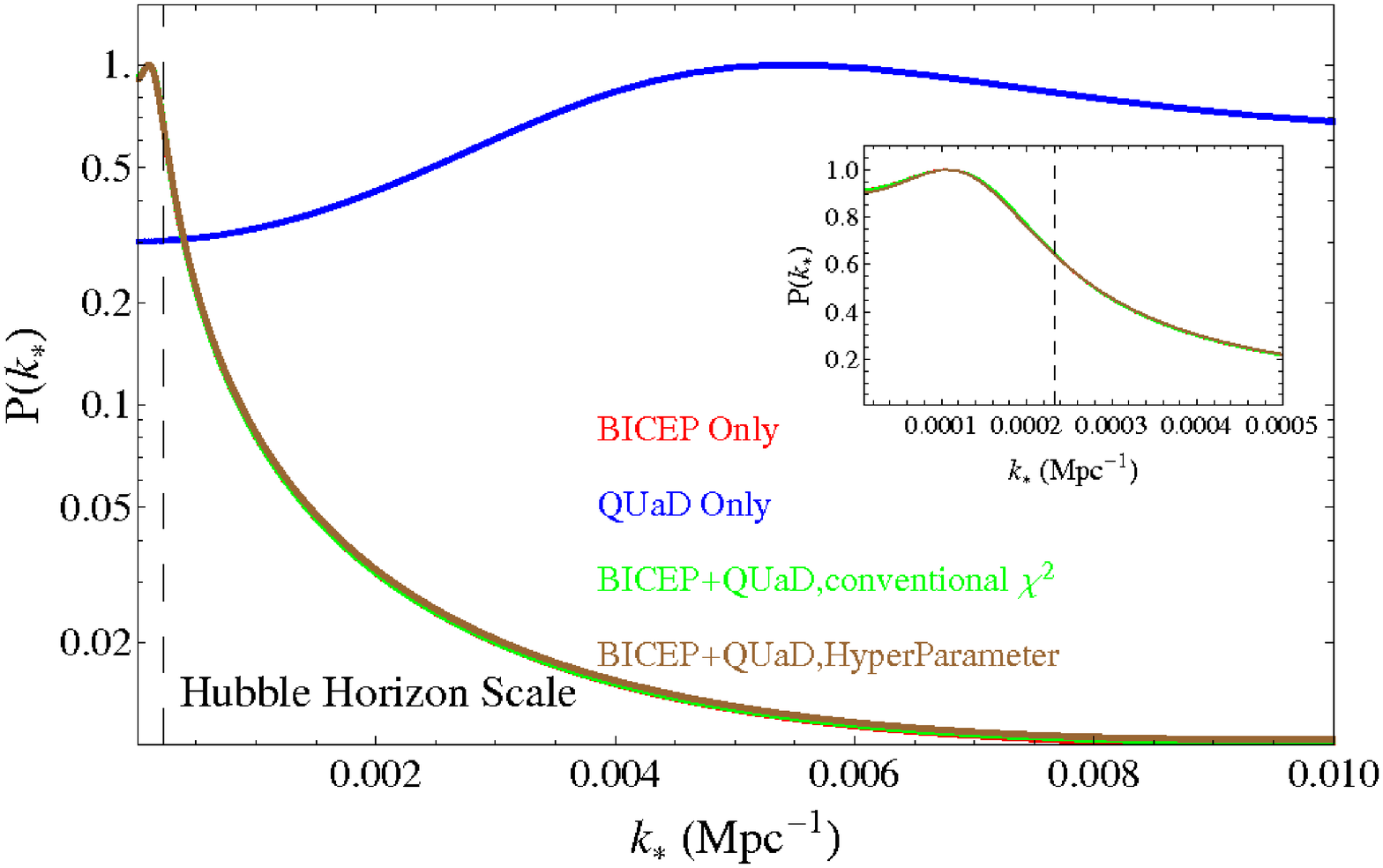}} \caption{Probability
distribution function (PDF) for the parameters in the LQG model. Left:
constraints on the mass parameter $\tilde{m}=m/\text{M}_{\text{pl}}$.
Right: constraints on $k_{*}$. The red, green, and brown curves
overlap with each other, indicating that the vast majority of the
constraining power for the combined data sets comes from the BICEP
data.} \label{loopconstraint}
\end{figure}

The current constraints on the parameters $k_*$ and $m$ are shown in
Fig.~\ref{loopconstraint}. In the left panel, we have marginalized
over the parameter $k_{*}$ and we plot the PDF for the mass parameter
$m$. The $1\sigma$ upper bound is $m \leq 1.36
\times 10^{-8} \text{M}_{\text{pl}}$. The detailed results are listed
in the second and third rows of Table \ref{tab1}. We note that the GUT
scale mass $m \simeq 10^{-7} \text{M}_{\text{pl}}$ is excluded at the
$2 \sigma$ level, but is still well within $3 \sigma$. As was the case
with the SFI models, the small-scale QUaD data is unable to constrain
the LQG models. Once again, the combined constraints are dominated by
the BICEP data as is clear from the figure and from the results listed
in Table 1.

In the right panel of Fig. \ref{loopconstraint}, we show the PDF for
$k_{*}$ (marginalized over the $m$ parameter). Once again the results
are dominated by BICEP. From this plot, wee see that there is a peak
in the PDF at $k_{*}=1.07 \times 10^{-4} \text{Mpc}^{-1}$. Although it
is not statistically significant, a detection of such a feature would
be an interesting result for the pre-inflationary bouncing behavior,
since the bounce of the primordial tensor power spectrum is
characterized by a non-zero $k_{*}$ as we have already discussed. We
further note that the peak value of $k_{*}$ is only slightly smaller
than the current Hubble wavenumber $k_{H}$, which would indicate that
modes which had the same length scale as the Hubble horizon at the
beginning of inflation have not evolved into the Hubble horizon
yet. In that case, the bump in the tensor power spectrum of LQG is a
super-horizon feature. However, we stress again that all of our
results are upper limits only and our formal constraint is $k_{*} <
2.43\times 10^{-4} \text{Mpc}^{-1}$ ($1\sigma$ CL).

\begin{figure}[tbp]
\centerline{\includegraphics[bb=0 0 540
354,width=4.2in]{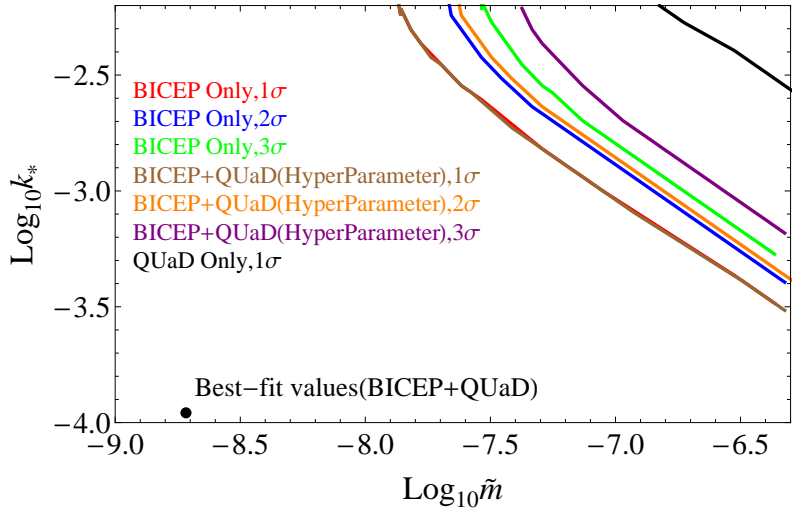}} \caption{2D constraints on the
  parameters $\tilde{m}=m/\text{M}_{\text{pl}}$ and
  $k_{*}$[$\text{Mpc}^{-1}$] of the LQG model. The red line (BICEP $1
  \sigma$ CL) overlaps with the brown line (BICEP+QUaD, $1 \sigma$ CL).}
\label{loopconstraint2}
\end{figure}
In Fig. \ref{loopconstraint2}, we plot the two-dimensional constraints
on the parameters $m$ and $k_{*}$ on a log scale. Clearly, the current
data is unable to provide strong constraints on the joint distribution
of these two parameters and can only provide upper limits. Once again,
as expected, the constraints are dominated by the BICEP data.


\subsection{Prospects for future experiments}
\label{fisher-loop}

\begin{figure}[tbp]
\centerline{\includegraphics[bb=0 0 616
412,width=3.4in]{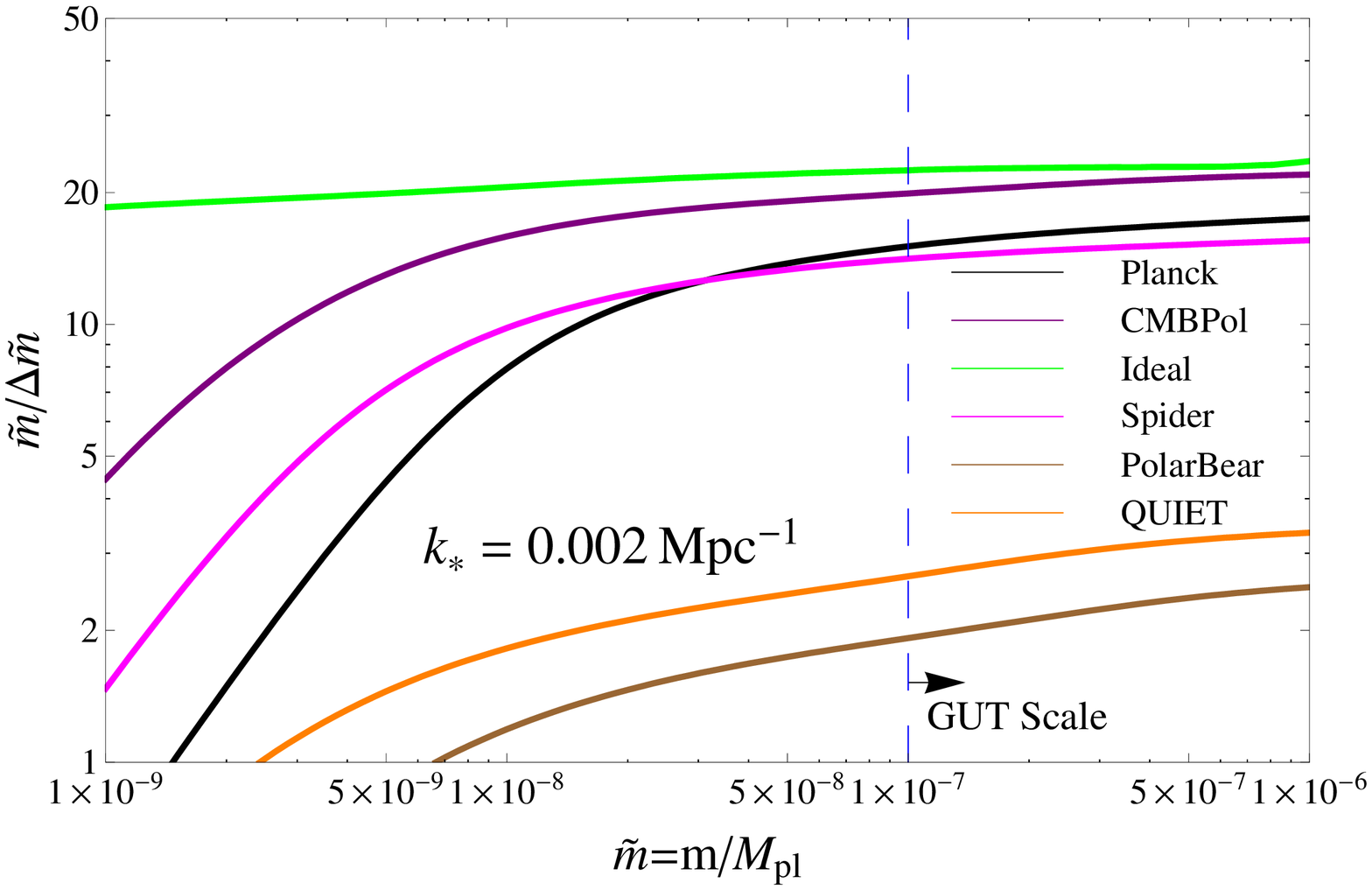}\includegraphics[bb=0 0 593
390,width=3.4in]{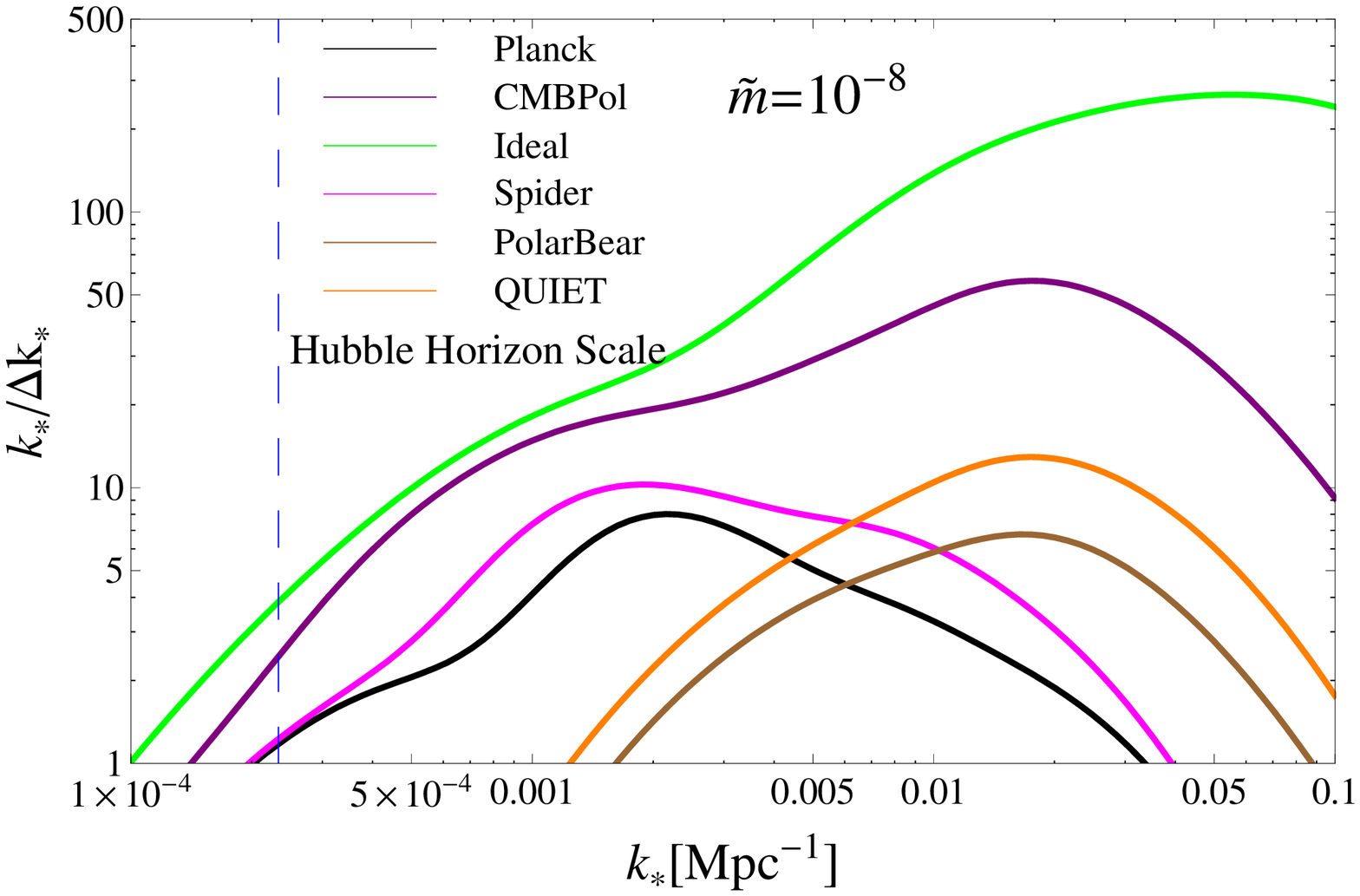}}
\caption{Predicted signal-to-noise ratios for the parameters in LQG
  for forthcoming and future experiments. Here $\tilde{m}$ is the mass
parameter, and $k_{*}$ is the position of the bump in the BB power
spectrum.}
\label{snratioloop}
\end{figure}
\begin{figure}[tbp]
\centerline{\includegraphics[bb=0 0 463
435,width=3.4in]{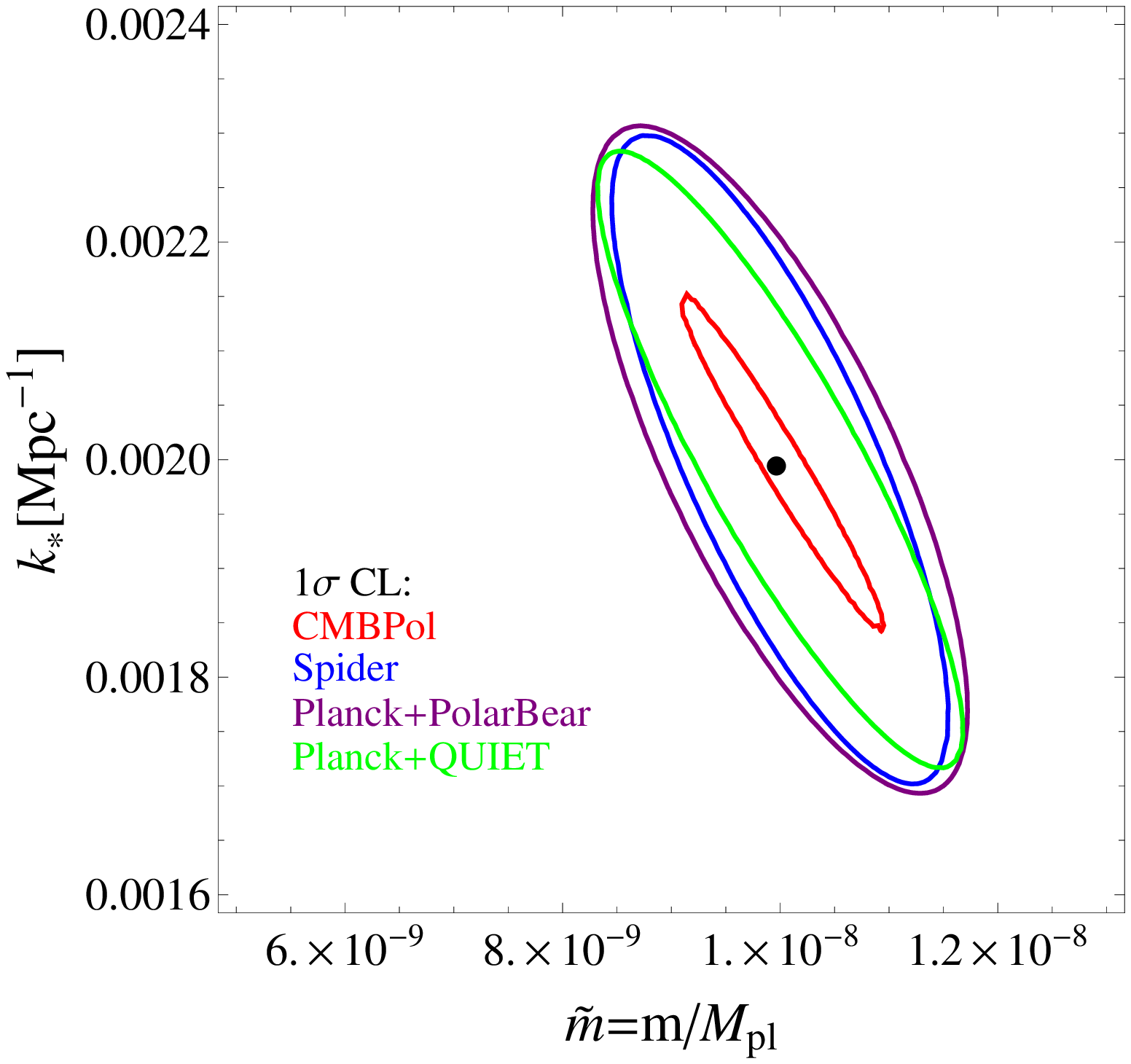}\includegraphics[bb=0 0 486
457,width=3.4in]{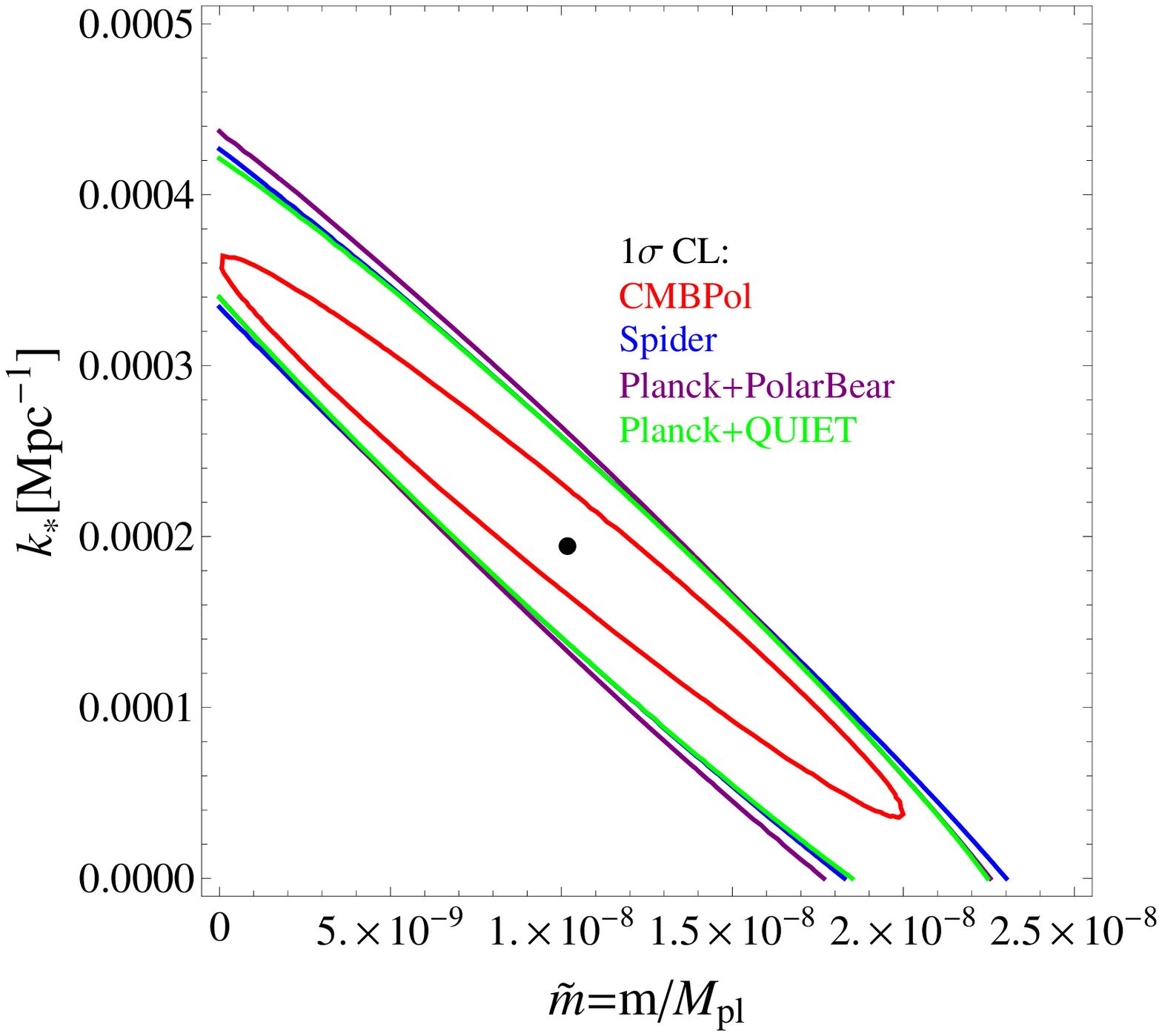}}
\caption{Forecasted 2D constraints for the LQG model. The input models
  are indicated with the black points. (Left: $\tilde{m}=10^{-8}$ and
$k_{*}=0.002 \text{Mpc}^{-1}$. Right: $\tilde{m}=10^{-8}$ and
$k_{*}=0.0002 \text{Mpc}^{-1}$).}
\label{loopcontours}
\end{figure}
We now investigate the prospects for detection of LQG signatures with
future CMB experiments by studying the projected constraints on the
two parameters $m$ and $k_*$. Once again, we use the Fisher matrix
approach described in Section~\ref{fishermatrix} and as in Section
\ref{fisher-sfi}, we fix the background parameters at their WMAP
7-year best-fit values \cite{Komatsu10}.

In Fig. \ref{snratioloop}, we plot the projected signal-to-noise
ratio for the parameters $m$ (left panel) and $k_*$ (right panel)
for forthcoming experiments. In the left panel, we plot the
signal-to-noise ratio $m/\Delta m$ as a function of $m$. For this
plot, we have kept $k_*$ fixed at $k_*=0.002$ Mpc$^{-1}$.  We find
that, if $m>10^{-7}$ M$_{\rm pl}$, i.e. the energy scale of
inflation is higher than the GUT scale, then Planck, Spider and
CMBPol could potentially detect the LQG signal at more than
$10\sigma$ due to their large sky coverage. In contrast, the
smaller scale experiments (PolarBear and QUIET) would only detect
the signal at the 2---3$\sigma$ level. For $m=10^{-8}$ M$_{\rm
pl}$, close to the upper bound of the current $1\sigma$ confidence
level, the large scale survey experiments (Planck, Spider and
CMBPol) can still detect the signal at more than $5\sigma$. For
this mass parameter and value of $k_*$, the ground-based
experiments, PolarBear and QUIET, would be insensitive to the
signal since the $B$-mode power spectrum from LQG falls off
extremely rapidly with increasing $l$. However, for larger values
of $k_*$, the peak of the LQG $C_l^{BB}$ power spectrum moves to
smaller scales. We therefore expect a general trend whereby the
large-scale experiments will be sensitive to models with small
values of $k_*$ and small-scale experiments will be sensitive to
models with larger values of $k_*$.

This is illustrated in the right-hand panel of Fig.~\ref{snratioloop}
where we plot forecasts for $k_*/\Delta k_*$ as a function of
$k_*$. For these results, we have fixed $m=10^{-8}$ M$_{\rm pl}$. We
find that the signal-to-noise ratio of $k_{*}$ does not monotonically
increase with increasing $k_*$ for any single experiment. Since this
parameter controls the angular scale at which the LQG $B$-mode signal
peaks, as we vary $k_*$ we move between the sensitivity ranges of
different experiments. For example at $k_* \approx 0.002$ Mpc$^{-1}$,
both Planck and Spider could detect the signal at more than $5\sigma$
whereas PolarBear and QUIET would achieve only a marginal
detection. Conversely, if $k_* \approx 0.02$ Mpc$^{-1}$, the reverse
is true: PolarBear and QUIET would make strong ($\gtrsim 5\sigma$)
detections while Planck and Spider would struggle to detect a signal.


In Fig. \ref{loopcontours}, we plot the two-dimensional constraints on
the parameters $m$ and $k_*$ for two typical models. The left panel
shows the forecasted constraints for a model with $k_*=0.002$ Mpc$^{-1}$
and the right panel shows the constraints for a model with
$k_*=0.0002$ Mpc$^{-1}$. In both cases, a fiducial value of
$m=10^{-8}$ M$_{\rm pl}$ was adopted. We find that the former case can
be well constrained by either Spider, CMBPol, Planck+PolarBear
or Planck+QUIET while the latter case can only be meaningfully
constrained by the CMBPol mission.

\section{Cosmic strings and their detection}
\label{cosmic-string}
\subsection{$B$-mode polarization from cosmic strings}
\label{Bmode-cosmic-string}  Cosmic strings have been proposed as
a possible source of the inhomogeneities in the Universe
\cite{cosmicstring}. Although current observations of the CMB
temperature and polarization power spectra suggest that it is
inflation, rather than cosmic strings, which is the main source of
the primordial density perturbations \cite{Komatsu10}, there is
still significant motivation to search for the signature of cosmic
strings from both theoretical and observational considerations.

Cosmic strings can be formed in several inflationary pictures, and
particularly in Brane inflation models
\cite{kklt,Kachru03,braneinflation}.  Brane inflation arises from
the framework of high dimensional string theory and is a further
important model for sourcing the dynamics of inflation. In this
model, the high dimensional Brane and anti-Brane collided and
annihilated and cosmic strings were produced at the end of the
inflationary epoch. If this scenario is correct, the resulting
strings would have made an observable imprint on the CMB sky by
way of the Kaiser-Stebbins effect \cite{kseffect}.

Although current observations suggest that inflation sources the
majority of the CMB anisotropy, one cannot rule out a significant (up
to $\sim 10\%$) contribution from cosmic strings
\cite{stringconstraint}. In this section, we will use CMB data to
constrain the level of cosmic strings. However, in contrast to other
works (see e.g.~\cite{stringconstraint}), we focus on the possible
detection of cosmic strings through the $B$-mode power spectrum
alone. $B$-mode polarization can only be generated in the early
Universe by vector and tensor perturbations, which provides a
complementary route for detecting cosmic strings. Although the
contribution of scalar perturbations from cosmic strings is
subdominant ($< 10\%$) compared to the contribution from SFI, the
contributions of vector and tensor perturbations from cosmic strings
may constitute a very significant fraction of the $B$-mode
polarization power on small angular scale (high multipoles)
\cite{Pogosian08}.

To predict the $C_{l}^{BB}$ power spectrum generated by cosmic
strings, one must understand the evolution of a cosmic string network
and it is important to know the characteristics of the scaling regime
which needs to be assumed in the numerical simulation. There are two
popular ways of making progress. One is to solve the Nambu equations
of motion for a string in an expanding universe and ignore the effects
of radiation backreaction (hereafter the Nambu-String model); the
other is to solve the equations of motion for the Abelian-Higgs (AH)
model, but to limit the dynamical range of the simulation. In this
work, we will consider only the Nambu-String model but we note that
the Abelian-Higgs model can be constrained in a similar way.

In order to calculate the $B$-mode power spectrum, including the
contributions of both vector and tensor perturbations, we use the
publicly available code CMBACT \cite{CMBACT,Pogosian99} to
generate a fiducial $C_{l}^{BB}$ for cosmic strings with the
tension of the strings set to $G\mu_{0}=10^{-7}$. Since the
amplitude of the $B$-mode power spectrum generated by cosmic
strings is simply proportional to the square of the cosmic string
tension, we can scale the fiducial spectrum to any other value for
the cosmic string tension using
\begin{equation}
C^{BB}_{l}=C_{l}^{BB, 0}\left( \frac{G\mu }{G\mu _{0}}\right)^{2},
\end{equation}%
where $C_{l}^{BB, 0}$ is the power spectrum normalized at $G\mu_{0}=10^{-7}$.

\subsection{Constraints from current data}
\label{constraints-string}
\begin{figure}[tbp]
\centerline{\includegraphics[bb=0 0 610
403,width=4.2in]{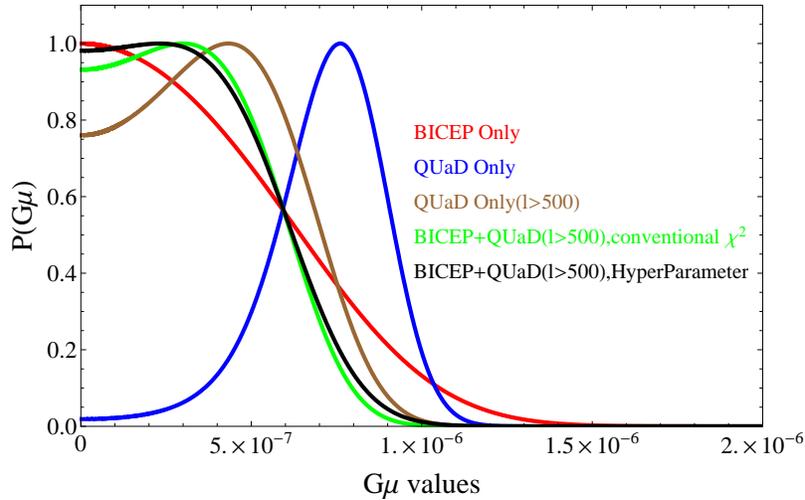}} \caption{Current constraints
from $B$-mode polarization on the cosmic string tension. To obtain the
constraints, we have fixed the wiggling parameter at
$\alpha=1.9$.} \label{stringconstraint}
\end{figure}
In Fig. \ref{stringconstraint}, we show the current constraints on
the cosmic string tension $G \mu$ from the BICEP and QUaD
$C_l^{BB}$ data. The upper bound from BICEP alone is $G \mu \leq
9.961 \times 10^{-7}$ ($2\sigma$ CL). Using the QUaD data alone
the result is $G \mu=7.60^{+2.63}_{-3.60} \times 10^{-7} \text{ }
(2 \sigma~{\rm CL})$, shown as the blue curve in
Fig.~\ref{stringconstraint}. Taken at face value, the QUaD result
represents a $2.8 \sigma$ detection of the cosmic string tension.
Referring back to Fig.~\ref{datacompare} and the discussion in
Section~\ref{constraints-sfi}, this result is likely coming from
the apparent excess of power seen in the QUaD data on scales $300
< l < 500$. To investigate further, we have repeated the analysis
with the $l < 500$ QUaD data removed. This results in the
likelihood function for $G\mu$ shown as the brown curve in Fig.
\ref{stringconstraint}. Excluding the $l < 500$ QUaD data shifts
the peak of the likelihood significantly towards a smaller value
(best-fit $G \mu=4.32 \times 10^{-7}$), and the constraint is now
consistent with zero at the $1 \sigma$ CL. The $2 \sigma$ upper
limit becomes $G\mu < 8.57 \times 10^{-7}$ (see Table \ref{tab1}
for the full set of results). The fact that the QUaD result
changes drastically when we remove the $l < 500$ measurements
suggests a problem with the $l < 500$ data. As in the case of the
SFI constraints presented earlier, we note that the shape of the
QUaD data at $l < 500$ is clearly inconsistent with the expected
$B$-mode signal for cosmic strings. Once again, we suspect that
the anomalous signal seen in the QUaD data between $l \approx 300$
and $l \approx 500$ is likely due to unquantified systematics. We
therefore consider the QUaD result restricted to $l > 500$ to be a
more robust constraint and consequently we quote this as our main
result.

We now examine the constraints obtained from combining the BICEP data
with the $l > 500$ QUaD data. Since the peaks of the two likelihood
functions do not overlap, we will carefully consider both the
conventional $\chi^2$ analysis and the hyper-parameter $\chi^2$. When
we use the conventional $\chi^2$, the peak of the combined likelihood
lies midway between the peaks of the two individual likelihoods (green
line in Fig.~\ref{stringconstraint}), and the resulting constraint is
$G\mu<7.72\times 10^{-7}$ ($2\sigma$ CL). If we use the
hyper-parameter $\chi^2$, the peak of the joint distribution moves
slightly further towards zero, and the best-fit is $G\mu<8.01\times
10^{-7}$ ($2\sigma$ CL). Since the conventional and hyper-parameter
approaches give very similar results, this suggests that the BICEP and
$l > 500$ QUaD data are mutually consistent which adds confidence to
the joint constraint.

A number of previous works have also attempted to constrain
cosmic strings through their imprint on CMB ($TT, TE, EE$),
large scale structure and gravitational waves data
\cite{constrain-bbn,constrain-pulsar,constrain-ligo,constrain-cs1,constrain-cs2,constrain-cs3}.
In a recent work \cite{stringconstraint}, constraints on the cosmic
string tension were obtained from a combination of CMB data (including
the WMAP 5-year, ACBAR, BOOMERGANG, CBI, QUAD and BIMA observations),
matter power spectrum data from the SDSS Luminous Red Galaxies sample,
and Big Bang Nucleosynthesis constraints on the baryon fraction from
measurements of deuterium at high redshift. They obtained a combined
upper limit of $G\mu<2.2\times 10^{-7}$ ($2\sigma$) in the
Nambu-String case.

As expected, this limit is much tighter than what we have obtained
from $B$-modes alone since current measurements of the $TT, TE$ and
$EE$ power spectra are much stronger than the $BB$ data that we have
used here. However, we note that in the analysis of
Ref. \cite{stringconstraint}, the main constraining power comes from
the CMB $TT$ and SDSS data and ultimately constraints from such data
will be limited by degeneracies with other parameters (most notably,
the spectral index $n_s$). As is the case for SFI models, the
advantage of using the $BB$ power spectrum to constrain the cosmic
string tension is that $C_{l}^{BB}$ is only very weakly dependent on
the other cosmological parameters, e.g. $n_s$. Our constraint is
therefore an independent check of other constraints obtained on $G\mu$
using different data and our result, although weaker, is consistent
with previous analyses. Cosmic string constraints from forthcoming CMB
polarization experiments will likely close the gap with other
techniques in terms of constraining power. We now turn to examining
the constraints achievable with these forthcoming experiments.

\subsection{Prospects for future experiments}
\label{fisher-string}


\begin{figure}[tbp]
\centerline{\includegraphics[bb=0 0 656
427,width=4.2in]{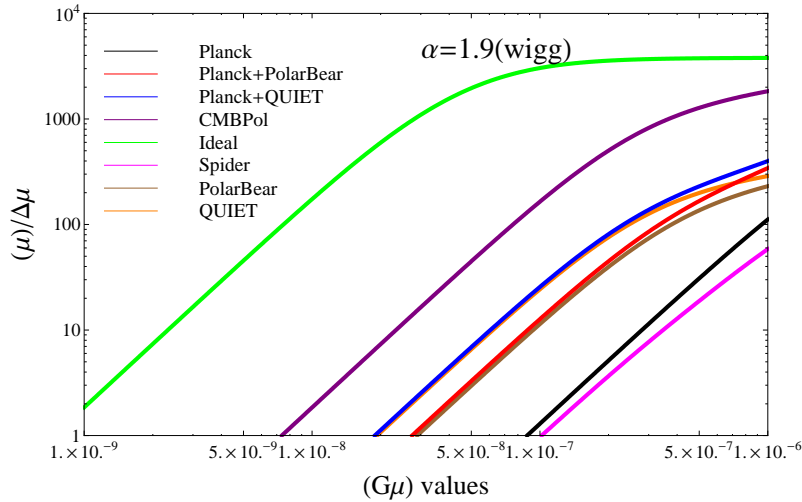}} \caption{Forecasts of the signal-to-noise
ratio for the cosmic string tension $G\mu$ potentially achievable with
future CMB polarization experiments.} \label{cosmic}
\end{figure}
In this subsection, we forecast the detectability of the cosmic string
tension $G\mu$ for future experiments. Once again, we use the Fisher
matrix formalism described in Section~\ref{fishermatrix}. In
Fig.~\ref{cosmic}, we plot the signal-to-noise ratio for a measurement
of $G\mu$ for the various experiments. As one might expect, the Planck
and Spider experiments are unable to tightly constrain the cosmic string
tension since neither experiment will produce sensitive $B$-mode
measurements on small scales where the string signal peaks.

For example, the Planck satellite can detect a cosmic string signal at
the $3\sigma$ level only if $G\mu \gtrsim 2\times10^{-7}$ is
satisfied. The ground-based experiments, PolarBear and QUIET, will
obviously be better at constraining $G\mu$ than Planck and
Spider. PolarBear should be able to detect $G\mu>5.0\times10^{-8}$ (at
the $3\sigma$ level) while QUIET should be able to detect
$G\mu>4.0\times10^{-8}$ (again at $3\sigma$). Adding Planck to either
of these experiments does not change the results significantly.
CMBPol is much more sensitive than the other experiments for all
fiducial $G\mu$ values, and its $3\sigma$ detection limit is $G\mu
\gtrsim 1.5\times10^{-8}$. Finally, we find that the ideal CMB
experiment can detect cosmic strings at the $3\sigma$ level if $G\mu
\gtrsim 1.5\times 10^{-9}$ is satisfied, which represents the
fundamental detection limit for CMB $B$-mode polarization experiments.

\section{Conclusion}
\label{conclusion} In this paper, we have examined
the observational signatures of three different models of the early
Universe related to the inflationary process. These three models are
in turn motivated by three different aspects of microscopic physics:
single field slow-roll inflation from effective field theory (SFI),
loop quantum cosmology from loop quantum gravity (LQG) and cosmic
strings from Brane inflation and/or high dimensional string theory. We
have discussed their potential observational signatures in the
$B$-mode polarization of the CMB, and we have constrained the
parameters of each model using the latest CMB polarization data from
the BICEP and QUaD experiments. Using a Fisher matrix formalism we
have forecasted the constraints achievable on these models using
future CMB polarization observations from a number of experiments
including Planck (space), PolarBear (ground), QUIET (ground), Spider
(balloon), CMBPol (space) and an idealized experiment.


We first discussed the SFI model. From the Lyth bound relation, we know
that $r \sim 0.01$ is an important bound for inflation reaching the
GUT scale, and the inflaton evolves over the trans-Planckian
region. The constraints we obtained from current $B$-mode measurements
are shown in Fig.~\ref{rconstrain}. Using the BICEP data alone, we
find $r=0.01^{+0.31}_{-0.26}$ ($1\sigma$ CL) in close agreement with the
BICEP team's own analysis \cite{bicep}. As expected, this constraint
does not change significantly on the addition of the small-scale QUaD
data.
Looking to the future, we find that the Planck satellite may be
able to detect $r\sim0.05$ (at the $3\sigma$ CL), while the
Spider, QUIET and PolarBear experiments all have the potential to
make a $5\sigma$ detection of $r\sim0.05$. The possible future
satellite mission, CMBPol could detect $r \sim 0.01$ at the
$20\sigma$ CL, and could even detect $r \sim 0.002$ at the $\sim
3\sigma$ CL. (All of these forecasts are for the case where the
tensor spectral index is held fixed at $n_t = 0$.) In summary, we
find that all future experiments can potentially constrain the
value of $r$ with sufficient sensitivity to allow the SFI model
from effective field theory to be tested in an interesting way.

We have also discussed the LQG model, which predicts a
pre-inflationary bouncing era. Before the bounce, the Universe was
contracting and dominated by vacuum energy. Its tensor power
spectrum is characterized by two parameters: $m$ and $k_{*}$,
where the mass parameter $m$ controls the magnitude of the
$B$-mode power spectrum, and $k_{*}$ controls the scale of the peak in
the $B$-mode spectrum. Our joint likelihood analysis using current
data yields $m < 6.16 \times 10^{-8} \, \text{M}_{\text{pl}}$ and
$k_*<7.05\times10^{-4}$ Mpc$^{-1}$ (both $2\sigma$ CL upper limits).
The PDF for the parameter $k_{*}$ exhibits a peak at position
$k_*=1.07\times 10^{-4}$ Mpc$^{-1}$. Although this peak is not
statistically significant, we note that were the value of $k_*$ to be
around this value, this would constitute evidence for a
pre-inflationary bounce.
We have also presented forecasts for constraining the LQG model using
future CMB experiments. We find that if the value of $k_*$ is as large
as $0.002$ Mpc$^{-1}$, a number of ongoing and future experiments
(Planck, Spider and CMBPol) could potentially detect the signal as
long as $m>2.5\times10^{-9}$ M$_{\rm pl}$. However, if the value of
$k_*$ were to be as low as $0.0002$ Mpc$^{-1}$ (as is mildly indicated
by current data), then the signature of LQG becomes quite difficult to
detect. We find that, for a typical choice of $m=10^{-8}$ M$_{\rm pl}$,
only CMBPol and the ideal experiment could detect the signal for such
a low value of $k_*$.

Finally, we have presented current and prospective constraints on
the cosmic string tension. We find that the BICEP and QUaD data
constrain the cosmic string tension to be $G\mu \leq 9.96 \times
10^{-7}$ and $G\mu \leq 8.57 \times 10^{-7}$ respectively (both
$2\sigma$ CL upper limits). The combined constraint is $G\mu<8.01
\times 10^{-7}$, which is weaker, but comparable to the constraints
from the CMB temperature anisotropy power spectrum. In terms of
forecasts for the future, we find that the high-resolution ground-based
experiments, PolarBear and QUIET, are more useful for constraining
$G\mu$ (as compared to e.g. Planck and Spider) since they are
much more sensitive to the $B$-mode power spectrum on small angular
scales. These two experiments could detect $G\mu \sim 10^{-7}$ at more
than $10\sigma$ CL. We also find that CMBPol can detect the signal of cosmic
strings if $G\mu>1.5 \times 10^{-8}$ at $3 \sigma$ CL, and the ideal CMB
experiment can detect the signal at the $3\sigma$ level even if the
tension of cosmic string is as low as $G\mu=1.5 \times 10^{-9}$.

Although the $B$-mode polarization derived constraints which we have
presented in this paper are currently only upper limits, they are
nevertheless already comparable to the equivalent constraints obtained
using a combination of all other cosmological data. In terms of
constraining the parameters of early Universe models, $B$-mode
polarization is clearly a very powerful tool and will likely overtake
other early Universe probes with the advent of the next generation of
CMB polarization experiments. In this paper, we have not directly
considered the issue of model selection. However, it is likely that,
in addition to constraining model parameters, future sensitive
$B$-mode observations will also allow us to distinguish between models
of the early Universe such as those considered in this paper.


\begin{acknowledgments}
The authors acknowledge useful communications and discussions with Richard
Battye, Daniel Baumann, Cynthia Chiang, Anthony Challinor, George
Efstathiou, Eanna Flanagan, Christopher Gordon, Michael Hobson, Deepak
Baskaran, Leonid Grishchuk and Mao Zeng. Yin-Zhe Ma thanks Trinity
College Cambridge and Cambridge Overseas Trusts for support.
\end{acknowledgments}

\appendix

\section{Instrumental Characteristics of CMB Experiments}
\label{instruments}
\begin{table}[tbp]
\begin{center}
\begin{tabular}{|c|c|c|c|c|c|c|c|}
\hline Band center~[GHz] & 30 & 44 & 70 & 100 & 143 & 217 & 353 \\
\hline FWHM~[arcmin] & 33 & 24 & 14 & $10.0$ & $7.1$ & $5.0$ &
$5.0$ \\ \hline
$N_{\mathrm{ins},l}^{BB}(i)$~[$10^{-6}\mu$K$^2$] & 2683 & 2753 & 2764 & $%
504$ & $279$ & $754$ & $6975$ \\ \hline
\multicolumn{1}{|c|}{$f_{\mathrm{sky}}$} & \multicolumn{7}{c|}{$0.65$} \\
\hline
\end{tabular}%
\caption{Instrumental parameters for the Planck satellite (space-based
experiment) \protect\cite{planck}. Here we have assumed 4 sky surveys
(28 months).} \label{tableplanck}
\end{center}
\end{table}
\begin{table}[tbp]
\begin{center}
\begin{tabular}{|c|c|c|c|}
\hline Band center ~[GHz] & 90 & 150 & 220 \\ \hline FWHM~[arcmin]
& $6.7$ & $4.0$ & $2.7$ \\ \hline
$N_{\mathrm{ins},l}^{BB}(i)$~[$10^{-6}\mu$K$^2$] & 5.2 & 4.3 & 44.0 \\
\hline
\multicolumn{1}{|c|}{$f_{\mathrm{sky}}$} & \multicolumn{3}{c|}{$0.024$} \\
\hline
\end{tabular}%
\caption{Instrumental parameters for the ground-based PolarBear
  experiment \protect\cite{polarbear}.}
\label{tablepolarbear}
\end{center}
\end{table}

\begin{table}[tbp]
\begin{center}
\begin{tabular}{|c|c|c|}
\hline Band center ~[GHz] & 40 & 90 \\ \hline FWHM~[arcmin] & $23$
& $10$ \\ \hline
$N_{\mathrm{ins},l}^{BB}(i)$~[$10^{-6}\mu$K$^2$] & 0.26 & 0.64
\\ \hline
\multicolumn{1}{|c|}{$f_{\mathrm{sky}}$} & \multicolumn{2}{c|}{$0.04$} \\
\hline
\end{tabular}%
\caption{Instrumental parameters for the ground-based QUIET
experiment
  \protect\cite{quiet}. Here we have assumed the phase-2 experiment.}
\label{tablequiet}
\end{center}
\end{table}

\begin{table}[tbp]
\begin{center}
\begin{tabular}{|c|c|c|c|c|}
\hline Band center ~[GHz] & 100 & 145 & 225 & 275 \\ \hline
FWHM~[arcmin] & $58$ & $40$ & 26 & 21 \\ \hline
$N_{\mathrm{ins},l}^{BB}(i)$~[$10^{-6}\mu$K$^2$] & 84.4 & 47.4
& 395 & 1170 \\ \hline
\multicolumn{1}{|c|}{$f_{\mathrm{sky}}$} & \multicolumn{4}{c|}{$0.5$} \\
\hline
\end{tabular}%
\caption{Instrumental parameters for the balloon-borne Spider
  experiment \protect\cite{spider}. Here we have assumed a 30-day LDB flight.}
\label{tablespider}
\end{center}
\end{table}

\begin{table}[tbp]
\begin{center}
\begin{tabular}{|c|c|c|c|c|c|c|c|}
\hline Band center~[GHz] & 30 & 45 & 70 & 100 & 150 & 220 & 340 \\
\hline FWHM~[arcmin] & 26 & 17 & 11 & $8$ & $5$ & $3.5$ & $2.3$ \\
\hline $N_{\mathrm{ins},l}^{BB}(i)$~[$10^{-6}\mu$K$^2$] & 31.21
& 5.79 & 1.48 & 0.89 & 0.83 & 1.95 & 39.46 \\ \hline
\multicolumn{1}{|c|}{$f_{\mathrm{sky}}$} & \multicolumn{7}{c|}{$0.8$} \\
\hline
\end{tabular}%
\caption{Instrumental parameters for the mid-cost (EPIC-2m) CMBPol
  satellite mission \protect\cite{cmbpol}.}
\label{tablecmbpol}
\end{center}
\end{table}

\begin{table}[tbp]
\begin{center}
\begin{tabular}{|c|c|c|}
\hline Parameter & ~~~{Synchrotron}~~~ & ~~~{Dust}~~~ \\ \hline
$A_{S,D}$ & $4.7\times 10^{-5}$~$\mu$K$^{2}$ & $1.2\times10^{-4}$~$\mu$K$%
^{2} $ \\ \hline $\nu_0$ & 30~GHz & 94~GHz \\ \hline $l_0$ &
350 & 900 \\ \hline $\alpha$ & -3 & 2.2 \\ \hline $\beta^{BB}$ &
-2.6 & -1.4 \\ \hline
\end{tabular}%
\caption{Assumptions for foregrounds parameters \cite{cmbpol} }
\label{tableforeground}
\end{center}
\end{table}
To calculate the total noise power spectrum $N_{l}^{BB}$ (Section
\ref{fishermatrix}), we require the experimental specifications
of each experiment, including the levels of both residual foreground
noise and instrumental noise. We list the instrumental noise for each
frequency channel the experiments we have considered in Table
\ref{tableplanck}-\ref{tablecmbpol}. The effective noise power
spectrum $N_{l }^{BB}$ is given by the optimal combination of the
channels \cite{cmbpol}
\begin{equation}
\lbrack N_{l }^{BB}]^{-2}=\sum_{i\geq j}\left[ (N_{\mathrm{fg},l
}^{BB}(i)+N_{\mathrm{ins},l }^{BB}(i))(N_{\mathrm{fg},l }^{BB}(j)+N_{%
\mathrm{ins},l }^{BB}(j))\frac{1}{2}(1+\delta _{ij})\right] ^{-1},
\label{totalnoise}
\end{equation}%
where $N_{\mathrm{ins},l }^{BB}(i)$ and $N_{\mathrm{fg},l
}^{BB}(i)$ are the instrumental and residual foreground noise
power spectra, respectively. Note that the noise power spectra
$N_{\mathrm{ins},l }^{BB}(i)$ listed in the tables do not include
the window function of the instrumental beam $\exp[l(l+1)\theta^2_{F}/(8\ln2)]$.

To model polarized foregrounds, we focus on diffuse synchrotron (S)
and dust (D) emission. The foreground contamination can be quantified
by the parameter $\sigma ^{\mathrm{fg}}$ which multiplies the power
spectra $C_{S,l }^{BB}(i)$, $C_{D,l }^{BB}(i)$ of the foreground
models. The smaller the value of $\sigma ^{\mathrm{fg}}$ the deeper
the foreground cleaning. Throughout this paper, we adopt $\sigma
^{\mathrm{fg}}=0.1$. The residual foreground noise is given by (see
\cite{cmbpol} for instance)
\begin{equation}
N_{\mathrm{fg},l }^{BB}(i)=\sum_{{f}=S,D}\left[ C_{{f},l
}^{BB}(i)\sigma ^{\mathrm{fg}}+\mathcal{N}_{{f},l }^{BB}(i)\right]
, \label{fore}
\end{equation}%
where $\mathcal{N}_{f,l }^{BB}(i)$ is the noise power spectrum
arising from the cleaning procedure itself in the presence of
instrumental noise.

Following \cite{foreground3,cmbpol,foreground2}, we model the scale
($l$) and frequency ($\nu_{i}$) dependence of the synchrotron and dust
emission as
\begin{equation}
C_{S,l }^{BB}(i)=A_{S}\left( \frac{\nu _{i}}{\nu _{0}}\right)
^{2\alpha _{S}}\left( \frac{l }{l _{0}}\right) ^{\beta _{S}}
\label{synchr}
\end{equation}%
and
\begin{equation}
C_{D,l }^{BB}(i)=p^{2}A_{D}\left( \frac{\nu _{i}}{\nu _{0}}\right)
^{2\alpha _{D}}\left( \frac{l }{l _{0}}\right) ^{\beta
_{D}^{BB}}\left[ \frac{e^{h\nu _{0}/kT}-1}{e^{h\nu
_{i}/kT}-1}\right] ^{2}.  \label{dust}
\end{equation}%
In Eq.~(\ref{dust}), $p$ is the dust polarization fraction, estimated to be $%
5\%$ \cite{foreground3}, and $T$ is the temperature of the dust
grains, assumed to be constant across the sky with $T=18$K
\cite{foreground3}. Other parameters in Eqs.~(\ref{synchr}),
(\ref{dust}) are specified in Table \ref{tableforeground} taken from
\cite{cmbpol}.

The noise term $\mathcal{N}_{f,l }^{BB}(i)$ ($f=S,D$) entering Eq.~(\ref%
{fore}) is calculated in \cite{foreground3,cmbpol}
\begin{equation}
\mathcal{N}_{f,l }^{BB}(i)=\frac{N_{\mathrm{ins},l }^{BB}(i)}{n_{%
\mathrm{chan}}(n_{\mathrm{chan}}-1)/4}\left( \frac{\nu _{i}}{\nu _{\mathrm{%
ref}}}\right) ^{2\alpha }.
\end{equation}%
Here, $n_{\mathrm{chan}}$ is the total number of frequency
channels used in making the foreground template map, and $\nu
_{\mathrm{ref}}$ is the
frequency of the reference channel. In the case of dust, $\nu _{\mathrm{%
ref}}$ is the highest frequency channel included in the template
making, while in the case of synchrotron, $\nu
_{\mathrm{ref}}$ is the lowest frequency channel. The value of
$\alpha $ is given in Table \ref{tableforeground} for different
foreground models. We note that the ground-based experiments
are insensitive to the largest angular scales, so when calculating
the Fisher matrix using Eq. (\ref{falphabeta}), we sum over the $l$
from 21 to 3000. In addition, for ground-based experiments, the
small scale fluctuations are not very sensitive to the residual
foreground noise, and we can also pick out relatively clean
patchs of sky where the foreground contamination is minimal.
Therefore, to forecast the results for PolarBear and QUIET, we have
not included a residual foreground noise term.

In addition to instrumental and residual foreground noise,
gravitational lensing converts $E$-mode polarization into $B$-modes on
small angular scales, contaminating the primordial $B$-mode signal
\cite{lensing}. The lensed $C_{l }^{BB}(\mathrm{lens})$ will also
contribute to the total noise power spectrum $N_{l }^{BB}$. The total
noise power spectrum therefore becomes
\begin{equation}
N_{l ,tot}^{BB}\equiv N_{l }^{BB}+C_{l
}^{BB}(\mathrm{lens}). \label{nltot}
\end{equation}

For the ideal case, we assume that there is no instrumental or
foreground noise, and that we can successfully de-lens the CMB
observations to a level of about $1/40$ of the lensing signal
\cite{seljak}. In this case, the total effective noise power spectrum
is
\begin{equation}
N_{l ,tot}^{BB}=1/40\times C_{l }^{BB}(\mathrm{lens}).
\end{equation}
Finally, we adopt $f_{\rm sky}=0.8$ for the ideal experiment, which is
the same as that used to model CMBPol.

\section{Statistics of the conventional $\chi^{2}$ and the hyper-parameter $\chi^{2}$}
\label{hyper-parameter} In this appendix, we first review the
basic results of conventional $\chi^{2}$ statistics and then generalize
the analysis to the hyper-parameter technique.

\subsection{$\chi^{2}$ statistics}
A conventional joint $\chi ^{2}$ analysis will minimize the following
combined $\chi ^{2}$
\begin{equation}
\chi _{tot}^{2}=\sum_{j}\chi _{j}^{2},
\end{equation}%
where each $\chi _{n}^{2}$ follows the chi-square distribution
\begin{equation}
f(\chi _{n}^{2})=\frac{1}{2^{\frac{n}{2}}\Gamma \left( \frac{n}{2}\right) }%
(\chi _{n}^{2})^{\frac{n}{2}-1}\exp (-\frac{1}{2}\chi _{n}^{2}).
\label{gamma}
\end{equation}%
It is easy to show that this chi-square distribution is properly
normalized, i.e.
\begin{equation}
\int_{0}^{\infty }f(\chi _{n}^{2})d\chi _{n}^{2}=1,
\end{equation}%
and the expectation value and the variance are
\begin{equation}
E(\chi _{n}^{2})=n,\text{ }V(\chi _{n}^{2})=2n.
\label{chi2-expectation}
\end{equation}%
Therefore, the minimum $\chi ^{2}$ value of a properly constrained
model should satisfy the following relation
\begin{equation}
1-\frac{\sqrt{V(n)}}{E(n)}  \leq \frac{\chi _{\min }^{2}}{E(\chi
_{n}^{2})}\leq 1+\frac{\sqrt{V(n)}}{E(n)}.
\end{equation}%
For the $\chi^{2}$ with order $n$, this is
\begin{equation}
1-\sqrt{\frac{2}{n}}\leq \frac{\chi _{\min }^{2}}{n}\leq 1+\sqrt{\frac{2}{n}}%
.
\end{equation}%
If the $\chi _{\min }^{2}/n$ $\geq 1+\sqrt{\frac{2}{n}}$, we can
say that the model does not
provide a good fit to the data, whereas if $\chi _{\min }^{2}/n$ $%
\leq 1-\sqrt{\frac{2}{n}}$, we say that the model overfits the
data, which may mean that the model has redundant free parameters.

If there are $m$ constraints on the $n$ random variables, then $\chi
_{n}^{2}$ still follows the chi-square distribution, but with order
$n-m$ \cite{Rileybook}
\begin{equation}
f(\chi _{n}^{2})=\frac{1}{2^{\frac{n-m}{2}}\Gamma \left( \frac{n-m}{2}%
\right) }(\chi _{n}^{2})^{\frac{n-m}{2}-1}\exp (-\frac{1}{2}\chi
_{n}^{2}). \label{gamma_constraints}
\end{equation}%
It is straightforward to verify, that the shape of the
distribution does not change, but the expectation value and the
variance are changed simply as $n\rightarrow n-m$.

\subsection{Hyper-parameter $\chi^{2}$}
The hyper-parameter approach to combining the constraints from
different data sets can be useful in the case where the different
data sets have different levels of systematics. To weight each data
set, one should multiply each $\chi^{2}$ by a free parameter,
\begin{equation}
\chi^{2}_{\text{hyper}}=\sum_{j}\alpha_{j}\chi^{2}_{j},
\end{equation}
where $\alpha_{j}$ is the weight parameter for each data set. One
can marginalize these weight parameters in a Bayesian analysis,
and in \cite{Lahav99}, the authors found that instead of minimizing
the combined $\chi ^{2},$ one should instead minimize the following
quantity
\begin{equation}
\chi _{\mathrm{hyper}}^{2}=\sum_{j}n_{j}\ln \chi _{j}^{2},
\end{equation}%
where $n_j$ is the number of degrees of freedom for each data set. If
we only consider one data set, the hyper-parameter statistic becomes%
\begin{equation}
\chi _{\mathrm{hyper}}^{2}=n\ln \chi _{n}^{2}.
\end{equation}%
From Eq. (\ref{gamma}), using the following transformation
\begin{equation}
f_{Y}=f_{X}\left\vert \frac{dY}{dX}\right\vert ,
\label{transform}
\end{equation}%
one can show that the distribution of the hyper-parameter statistic is
given by%
\begin{equation}
g(\chi _{\mathrm{hyper}}^{2})=\frac{1}{n\cdot
2^{\frac{n}{2}}\Gamma \left(
\frac{n}{2}\right) }\exp (\frac{1}{2}\chi _{\mathrm{hyper}}^{2})\exp (-\frac{%
1}{2}\exp (\frac{1}{n}\chi _{\mathrm{hyper}}^{2})),
\label{hyper-dis}
\end{equation}%
which has already been properly normalized. (When calculating the
integral, one should use the transformation $\exp (\frac{x}{n})=y$).
One can then show that the expectation value and variance of the
Hyper-parameter distribution is
\begin{equation}
E(\chi _{\mathrm{hyper}}^{2})=n\left( \ln 2+\psi _{0}\left( \frac{n}{2}%
\right) \right) ,\text{ }V(\chi _{\mathrm{hyper}}^{2})=n^{2}\psi
_{1}\left( \frac{n}{2}\right) ,
\end{equation}%
where $\psi _{n}(x)$ is the ``digamma function" defined as derivatives
of the log Gamma function
\begin{equation}
\psi _{n}(x)=\frac{d^{n+1}}{dx^{n+1}}\ln \Gamma (x).
\end{equation}

If there are $m$ constraints on the $n$ random variables (e.g. $m$
parameters), then the distribution of the conventional $\chi _{n}^{2}$
follows Eq. (\ref{gamma_constraints}). It is then easy to show that
the form of the distribution is unchanged only if the $\chi
_{\mathrm{hyper}}^{2}$ is defined as
\begin{equation}
\chi _{\mathrm{hyper}}^{2}=(n-m)\ln \chi _{n}^{2}.
\end{equation}

\textbf{Proof }From Eqs. (\ref{gamma_constraints}) and
(\ref{transform}), one can derive the following PDF
\begin{eqnarray}
g(\chi _{\mathrm{hyper}}^{2}) &=&f(\chi _{n}^{2})\frac{d\chi
_{n}^{2}}{d\chi
_{\mathrm{hyper}}^{2}}  \nonumber \\
&=&f\left( \exp \left( \frac{\chi
_{\mathrm{hyper}}^{2}}{n-m}\right) \right) \frac{1}{n-m}\exp
\left( \frac{\chi _{\mathrm{hyper}}^{2}}{n-m}\right)
\nonumber \\
&=&\frac{1}{n-m}\exp \left( \frac{\chi
_{\mathrm{hyper}}^{2}}{n-m}\right)
\frac{1}{2^{\frac{n-m}{2}}\Gamma \left( \frac{n-m}{2}\right) }  \nonumber \\
&&\times \left[ \exp \left( \frac{\chi _{\mathrm{hyper}}^{2}}{n-m}\right) %
\right] ^{\frac{n-m}{2}-1}\exp \left( -\frac{1}{2}\exp \left( \frac{\chi _{%
\mathrm{hyper}}^{2}}{n-m}\right) \right)   \nonumber \\
&=&\frac{1}{n-m}\frac{1}{2^{\frac{n-m}{2}}\Gamma \left(
\frac{n-m}{2}\right)
}\exp \left( \frac{\chi _{\mathrm{hyper}}^{2}}{2}\right) \exp \left( -\frac{1%
}{2}\exp \left( \frac{\chi _{\mathrm{hyper}}^{2}}{n-m}\right)
\right) .
\end{eqnarray}%
Therefore, in the case of $m$ constraints, the distribution keeps its
form, and the expectation value and
variance become%
\begin{equation}
E(\chi _{\mathrm{hyper}}^{2})=(n-m)\left( \ln 2+\psi _{0}\left( \frac{n-m}{2}%
\right) \right) ,\text{ }V(\chi
_{\mathrm{hyper}}^{2})=(n-m)^{2}\psi _{1}\left(
\frac{n-m}{2}\right) .  \label{expectation-hyper}
\end{equation}%
Thus, to ensure that the form of the
$\chi^{2}_{\text{hyper}}$ distribution function is unchanged, the
$\chi^{2}_{\text{hyper}}$ needs to be defined as
\begin{equation}
\chi _{\mathrm{hyper}}^{2}=\sum_{j} n_{j}\ln \chi_{n_{j}}^{2},
\label{hyper-chi2}
\end{equation}
where $n_{j}=n_{\text{data}}-m$ is the number of degree of freedom.

The hyper-parameter approach is an objective way to weight each data
set when producing joint constraints. The value of the weight is
simply the value of the effective hyper-parameter, which is
defined as \cite{Lahav99}
\begin{equation}
\alpha_{\text{A}}=\frac{n_{\text{A}}}{\chi^2_{\text{A}}}
,\label{weight}
\end{equation}
where A specifies a particular data set. Therefore, the larger the
value of $\alpha$, the larger the weight that the particular data
set takes (see Table \ref{tab3}).

\end{document}